\documentclass[11pt]{article}
\usepackage{pifont}

\usepackage{dsfont}
\usepackage{tikz}

\usepackage{amsmath,amssymb,graphicx}
\usepackage{hyperref}
\usepackage{stackengine}

\usepackage[nosort]{cite}
\usepackage[Symbol]{upgreek}%for \upmu
\usepackage{amsmath,amssymb,graphicx}

\usepackage{hyperref}
\usepackage{multicol,color,longtable}
\definecolor{darkred}{rgb}{0.65,0.15,0}
\hypersetup{pdfborder={0 0 0},colorlinks=true,urlcolor=blue,citecolor=blue,linkcolor=darkred,linktocpage=true}

\usepackage[T1]{fontenc}
\usepackage{hyphenat}
\usepackage{amsfonts}
\usepackage{mathrsfs}
\usepackage{mathdots}

\usepackage{array}
\newcolumntype{P}[1]{>{\centering\arraybackslash}p{#1}}
\usepackage{lscape}
\usepackage{multirow}

\usepackage{empheq}

\newcommand{\be}{\begin{equation}}
\newcommand{\ee}{\end{equation}}
\newcommand{\bea}{\setlength\arraycolsep{2pt} \begin{eqnarray}}
\newcommand{\eea}{\end{eqnarray}}

\def\<{\langle}
\def\>{\rangle}

\newcommand{\eprint}[1]{{\href{http://arxiv.org/abs/#1}{[\texttt{#1}]}}}
\newcommand{\eprintN}[1]{{\href{http://arxiv.org/abs/#1}{[\texttt{#1 [hep-th]}]}}}

\newcommand{\doi}[2]{{\hypersetup{urlcolor=darkred}\href{http://dx.doi.org/#2}{#1}\hypersetup{urlcolor=blue}}}

\newcommand{\CR}{\nonumber \\*}

\makeatletter

\@addtoreset{equation}{section}
\makeatother

\def\i{\iota}						\def\m{\mu}

\newcommand{\Cong}[2]{\Gamma^{\scalebox{0.6}{#1}_{\scalebox{0.5}{#2}}}}

\newcommand{\sLambda}{I \hspace{-0.8mm}I}

\def\tr{{\rm tr}}

\begin{document}

\begin{flushright} CPHT-RR063.102023 \end{flushright} 
 \vspace{8mm}

\begin{center}

{\LARGE \bf \sc Non-geometric BPS branes on T-folds}

\vspace{6mm}
\normalsize
{\large  Massimo Bianchi${}^{1}$, Guillaume Bossard${}^{2}$}

\vspace{10mm}

${}^1${\it Dipartimento di Fisica, Universit\`a di Roma “Tor Vergata” \& Sezione INFN Roma2,\\
Via della Ricerca Scientifica 1, 00133, Roma, Italy}\\
${}^2${\it Centre de Physique Th\'eorique, CNRS,  Institut Polytechnique de Paris\\
91128 Palaiseau cedex, France}
\vskip 1 em
\vspace{20mm}

\hrule

\vspace{5mm}

 \begin{tabular}{p{14cm}}

We give a detailed (microscopic) description of the geometric and non-geometric fundamental branes and their bound states in Type II superstring compactifications preserving ${\cal N}=6$ supersymmetry. We consider general boundary states that couple to the twisted sector and compute the relevant annulus amplitudes. We check consistency of the construction by relating the ‘transverse’ channel, corresponding to closed-string `tree-level' exchange, with the ‘direct’ open-string loop channel. Focussing on the Type IIA frame, we show that D0-D4 have the expected tension for a geometric brane, while the non-geometric D2-D6 boundary states have a tension equal to $1/\sqrt{K}$  the one of a geometric brane for the $\mathds{Z}_K$ orbifold. This is consistent with Fricke T-duality of the $\mathcal{N}=6$ model. 
\end{tabular}

\vspace{6mm}
\hrule
\end{center}

\vspace{5mm} 

\thispagestyle{empty}

\newpage

\setcounter{page}{1}

\setcounter{tocdepth}{2}
\tableofcontents

%%%%%%%%%%%%%%%%%%%%%%%%
%%%%%%%%%%%%%%%%%%%%%%%%
%%%%%%%%%%%%%%%%%%%%%%%%
%%%%%%%%%%%%%%%%%%%%%%%%
\section{Introduction}

A large, probably the dominant,  part of string and brane configurations, ranging from vacuum configurations to black hole micro-states, may not admit a geometric description \cite{Hull:2004in, Hull:1994ys}. The simplest possibility is represented by configurations dubbed ``T-folds'' that require T-duality transformations in order to relate different locally geometric patches. Such vacua can be defined in perturbative string theory as asymmetric orbifolds  \cite{Narain:1986qm,Ferrara:1989nm,Narain:1990mw,Anastasopoulos:2009kj, Bianchi:2012xz}. Genuinely non-perturbative constructions, involving S-duality or U-duality transformations, are also conceivable \cite{Braun:2013yla, Candelas:2014jma, Candelas:2014kma, Garcia-Etxebarria:2015wns} and have attracted some attention recently \cite{Heckman:2020svr, Giacomelli:2023qyc}. 

Geometric branes can be defined on T-folds from the orbifold projection of branes of the parent theory that are invariant under the asymmetric orbifold action. Such branes never couple to the twisted sector of the closed string theory. A first microscopic description of D-branes in a trivial asymmetric orbifold was proposed in  \cite{Brunner:1999fj} for the $\mathds{Z}_2$ orbifold reflection on $T^4$ preserving $\mathcal{N}=8$ supersymmetry. The fundamental 1/2 BPS branes were constructed and couple to the twisted sector. This analysed was extended to $\mathds{Z}_2$ orbifolds of the bosonic string in \cite{Gaberdiel:2002jr,Kawai:2007qd}. The construction of these  boundary states strongly relies on the enhanced affine Kac--Moody symmetry one has at specific symmetric points of the Narain moduli space. It is desirable to identify which branes exist at generic points in moduli space to understand the non-perturbative dynamics of string theory. 

In order to get more control on the expected set of branes, it is useful to consider BPS branes in a supersymmetric theory. The largest is the amount of supersymmetry the more stringent are the consistency constraints. For this reason one can start by focussing on the largest non-maximal supersymmetry, i.e. 24 supercharges, that results from asymmetric orbifolds  of  Type II superstrings\cite{Ferrara:1989nm}. These backgrounds can be interpreted as a T-folds, with a $T^4$ fibered over $T^2$ with transition functions that are T-dualities.

Aim of the present investigation is to give a detailed (microscopic) description of the geometric and non-geometric fundamental BPS branes and their bound states in this class of backgrounds. This line of investigation was initiated in \cite{Bianchi:2008cj, Bianchi:2010aw} and recently revived in \cite{ Bianchi:2022tbr} in connection with higher derivative terms in the low-energy effective action for Type II superstrings with ${\cal N}=6$ supersymmetry. In particular,  in \cite{Bianchi:2022tbr}, it has been shown that only $\mathds{Z}_2$ and $\mathds{Z}_3$ orbifolds give rise to consistent vacuum configurations, thus ruling out $\mathds{Z}_4$ or higher order abelian groups.

We will derive the relevant annulus amplitudes both in the `transverse' channel, corresponding to closed-string `tree-level' exchange between (different) branes, and in the `direct' channel, corresponding one-loop amplitudes, coding the spectrum of  open-string excitations. 

To this end we will first briefly review the construction of the consistent  ${\cal N}=6$ Type II theories and write the relevant torus partition function, expressed in terms of  (super-)characters, in Section 2. For definiteness we work in the Type IIA frame, whereby the relevant branes are bound states of D0, D2, D4 and D6.

Focussing first on the $\mathds{Z}_2$ case, we identify and describe geometric and non-geometric fundamental branes in Section 3. We  consider general boundary states that couple to the twisted sector and compute their annulus partition function. In Section 4, we then show that geometric D2-D4-D6 branes can be viewed as bound-states of fundamental (non-geometric) branes.  In Section 5 we pass to consider the $\mathds{Z}_3$ asymmetric orbifold and describe   the geometric D0-D4 brane as well as the non-geometric D2-D6 brane. Our conclusions and an outlook are contained in Section 6. In an Appendix we carefully determine the allowed R-R charges and discuss the issue of rank reduction due to the presence of a discrete (non-dynamical) NS-NS antisymmetric tensor \cite{Bianchi:1990yu, Bianchi:1990tb}.

%%%%%%%%%%%%%%%%%%%%%%%% %%%%%%%%%%%%%%%%%%%%%%%%
\section{The torus partition function}
We consider type II string theory on a vacuum preserving $\mathcal{N}=6$ supersymmetry. The theory is obtained as an asymmetric orbifold that combines a  $\mathds{Z}_K$ rotation acting on the left-moving fields along $T^4$  as \cite{Anastasopoulos:2009kj, Bianchi:2012xz}
\be \tau_L ( X_L^i + i X_L^{i+2})   = e^{ \frac{2\pi i}{K}}  ( X_L^i + i X_L^{i+2})  \; , \qquad  \tau_L ( \psi_L^i + i \psi_L^{i+2})   = e^{ \frac{2\pi i}{K}} ( \psi_L^i + i \psi_L^{i+2}) \; , \ee
for $i=6,7$, as well as a shift on the circle $S^1$ of radius  $R_5$ acting on the bosonic coordinate as
\be \sigma X^5 = X^5  + \frac{2\pi}{K}  R_5 \; . \ee
 This is the simplest configuration of a T-fold  \cite{Hull:2004in}, for which one has locally a $T^4$ fibered over $S^1$, with the transition functions that are T-dualities of the $T^4$ world-sheet fields. 
 
 It was found in \cite{Bianchi:2022tbr} that this orbifold is consistent for $K=2$ and $3$ only. The construction of geometric D-branes was initiated for these models in   \cite{Bianchi:2008cj}. In this paper we  consider more general boundary states that couple to the twisted sector and compute their annulus partition function.

For this purpose we first give some notations and recall the torus partition function. We will start with the $\mathds{Z}_2$ orbifold and will discuss the $\mathds{Z}_3$ orbifold in a later section.

The $D_4$ lattice admits a consistent asymmetric $\mathds{Z}_2$ action, while a consistent asymmetric $\mathds{Z}_3$ action requires the lattice $A_2\oplus A_2$. The $D_4$ root lattice is isomorphic to the ring $H(2)$ of Hurwitz quaternions while the $A_2\oplus A_2$ root lattice is isomorphic to the ring $H(3)$ of Eisenstein quaternions \cite{ Bianchi:2022tbr}.\footnote{Denoting by $e_i$ the quaternions imaginary units, satisfying $e_i e_j = -\delta_{ij} + \varepsilon_{ij}{}^k e_k$, the ring of Hurwitz quaternions is generated over $\mathds{Z}$ by $1$, $e_1$, $e_2$ and 
$ -{\frac12}(1+e_1+e_2+e_3)$, while the ring of Eisenstein quaternions is generated over $\mathds{Z}$ by $1$, $-\frac{1}{2} + \frac{\sqrt{3}}{2}e_3$, $e_1$ and $-\frac{1}{2} e_1+ \frac{\sqrt{3}}{2}e_2$.}

The type II torus partition functions can  be written in terms of level 1 affine $SO(2n)$ characters \eqref{SO2nChar} as \cite{Bianchi:1990yu, Bianchi:1990tb}
\be
\mathcal{Z}^{T^d}_{{\rm IIA}} =  \frac{V_8-S_8}{\eta^8} \frac{\bar{V}_8-\bar{C}_8}{\bar{\eta}^8} \Lambda_{\sLambda_{d,d}}  \quad , \quad \mathcal{Z}^{T^d}_{{\rm IIB}} =\frac{V_8-S_8}{\eta^8}\frac{\bar{V}_8-\bar{S}_8}{\bar{\eta}^8}  \Lambda_{\sLambda_{d,d}}  \; ,
\ee
where we have not  included the (divergent) integral over the bosonic (center-of-mass and momentum) zero modes, while
\be \Lambda_{\sLambda_{d,d}} = \sum_{Q \in \sLambda_{d,d}} e^{ i \pi \tau p_L(Q)^2  -  i \pi \bar \tau p_R(Q)^2} \; , \ee
denotes the Narain partition function for the Lorentzian lattice $\sLambda_{d,d}$ and $p_L$ and $p_R$ are the dimensionless momenta.

It will also be  useful to introduce the $\mathcal{N}=2$ supersymmetric characters
\be    \mathcal{V}  = V_4 \chi_0 - S_4 \chi_\frac12 \; , \qquad \mathcal{H}  = O_4 \chi_{\frac12} - C_4 \chi_0 \; , \ee 
with $\chi_j$ the $SU(2)_1$ character of spin $j$ \eqref{SU2Char}. The notation $\mathcal{V},\mathcal{H}$ reminds us that $\mathcal{V}$ has the field content of a vector multiplet at the lowest (zero) mass level, while $\mathcal{H}$ has the field content of a `half' hyper-multiplet at the lowest (zero) mass level. The $\mathcal{N}=4$ supersymmetric character decomposes as 
\be \mathcal{Q}= V_8 - S_8 =   \mathcal{V} \chi_0 + \mathcal{H} \chi_{\frac12} \; . \label{QgivesVH}  \ee
By a slight abuse of language, we shall write both the type IIA right-moving fermion sector ${\bar{\cal Q}}_A = \bar{V}_8 - \bar{C}_8$ and the type IIB one ${\bar{\cal Q}}_B  = \bar{V}_8 - \bar{S}_8 $ as
\be \bar{\mathcal{Q}} = \overline{\mathcal{V} \chi_0 + \mathcal{H} \chi_{\frac12}}\; .  \ee
using that the two are identical up to a permutation of the two $SU(2)_1$ characters in $S_4$ and $C_4$ \eqref{OVCS4}. This permutation will be implicit in the definition of the Ishibashi states in type IIA.

The $\mathds{Z}_2$ orbifold is defined at the $D_4$ symmetric point  in Narain moduli space, where the $T^4$ metric and Kalb--Ramond fields are defined as in  \eqref{GBZ2} and the cross components of the metric and Kalb--Ramond fields  with one leg on $T^2$ and one leg on $T^4$ are equal. For simplicity, we shall assume that the cross components all vanish, such that the Narain partition function factorises as
\be \frac{\Lambda_{\sLambda_{6,6}}}{|\eta|^{12}} = \frac{\Lambda_{\sLambda_{2,2}}}{|\eta|^4}  \bigl( |O_{8}|^2 + |V_8|^2 + |S_8|^2 + |C_8|^2 \bigr) \; . \ee
We define 
\be \Lambda_{\sLambda_{2,2}}[^s_r] = \sum_{\substack{m_1,n_1,m_2\in \mathds{Z}\\ n_2 \in \mathds{Z}+\frac{s}{K}}} e^{ \frac{2\pi i}{K} r m_2} e^{ \frac{i \pi \tau }{2 T_2 U_2} | U m_1 -m_2 + T (n_1+U n_2)|^2 -  \frac{i \pi \bar \tau }{2 T_2 U_2} | U m_1 -m_2 + \bar T (n_1+U n_2)|^2} \ee
with $T = T_1 + i T_2$ the K\"{a}hler structure and $U=U_1 + i U_2$ the complex structure on $T^2$.

Then the type II $\mathds{Z}_2$ orbifold partition function can be written as  \cite{Bianchi:1990yu, Bianchi:1990tb}
\bea \mathcal{Z}^{T^6/\mathds{Z}_2}_{{\rm II}} &=& \frac12 (   \mathcal{V} \chi_0 + \mathcal{H} \chi_{\frac12}  )\bar{\mathcal{Q}}  \frac{\Lambda_{\sLambda_{2,2}}[^0_0]}{|\eta^4|^2} \bigl( |O_8|^2  +|V_8|^2 +|S_8|^2 +  |C_8|^2   \bigr) \\  &&  + \frac12(   \mathcal{V} \chi_0 - \mathcal{H} \chi_{\frac12}  )\bar{\mathcal{Q}}  \frac{\Lambda_{\sLambda_{2,2}}[^0_1]}{|\eta^4|^2}   \frac{4\eta^2 }{\vartheta_2(0)^2} \bar O_8\CR
&& + \frac12 (   \mathcal{V} \chi_{\frac12} + \mathcal{H} \chi_{0}  )\bar{\mathcal{Q}}  \frac{\Lambda_{\sLambda_{2,2}}[^1_0]}{|\eta^4|^2}  \frac{2\eta^2 }{\vartheta_4(0)^2} \bigl( \bar O_8  + \bar V_8 + \bar S_8 + \bar C_8\bigr)  
\CR
&& + \frac12 (   \mathcal{V} \chi_{\frac12} - \mathcal{H} \chi_{0}  )\bar{\mathcal{Q}}  \frac{\Lambda_{\sLambda_{2,2}}[^1_1]}{|\eta^4|^2}  \frac{2\eta^2 }{\vartheta_3(0)^2} \bigl( \bar O_8  - \bar V_8 - \bar S_8 - \bar C_8\bigr)    \; . \nonumber \eea
Although this partition function vanishes, thanks to supersymmetry, it conveniently exhibits the spectrum of the theory. To read the spectrum, we need to distinguish the cases in which the left-momentum on $T^4$ vanishes or not. If it does not vanish, the two states with opposite left-momenta are identified, with a sign depending on the left-fermion $SU(2)_1$ states and the parity of $m_2$. When the left-momentum along $T^4$ vanishes, the right-momentum must be in the $D_4$ lattice in the untwisted sector. Then the parity of the $SU(2)_1$ fermion state, the parity of the torus momentum $m_2$ and the parity of the mode number of the $T^4$ bosons must be the same.

At the symmetric point one can also make use of the level one affine $SO(8)$ algebra to fermionise the $T^4$ bosons. The orbifold action only preserves the $SO(4) \times SO(4)$ symmetry, so we decompose the Hilbert space into four $SU(2)_1$ modules. Then the action of the orbifold is defined as $(-1)^{m_2}  (-1)^{2 j_\psi} (-1)^{2 j_o}$  for $j_\psi$ the  $SU(2)_1$ spin in \eqref{QgivesVH} and $j_o$ the first $SU(2)_1$ of the $T^4$ bosons, so that
\bea\mathcal{Z}^{T^6/\mathds{Z}_2}_{{\rm II}}  &=& \frac12 \bigl(   \mathcal{V} \chi_0 + \mathcal{H} \chi_{\frac12} \bigr)\bar{\mathcal{Q}}  \frac{\Lambda_{\sLambda_{2,2}}[^0_0]}{|\eta^4|^2} \Bigl[\bigl( \chi_0^4 + \chi_{\frac12}^4\bigr)  \bar O_8  + \bigl( \chi_0^2 \chi_{\frac12}^2 +\chi_{\frac12}^2 \chi_{0}^2\bigr) \bar V_8 \\ 
&&\hspace{40mm} +  \bigl( \chi_0 \chi_{\frac12} \chi_0 \chi_{\frac12} +\chi_{\frac12}  \chi_0 \chi_{\frac12} \chi_0  \bigr) \bar S_8+  \bigl( \chi_0 \chi_{\frac12}^2 \chi_0 +\chi_{\frac12}  \chi_0^2 \chi_{\frac12}   \bigr) \bar C_8  \Bigr] \CR
%%%%%%%%%%%%%%%%%%%%%%%
&&  + \frac12 \bigl(   \mathcal{V} \chi_0 - \mathcal{H} \chi_{\frac12} \bigr)\bar{\mathcal{Q}} \frac{\Lambda_{\sLambda_{2,2}}[^0_1]}{|\eta^4|^2} \Bigl[  \bigl( \chi_0^4 - \chi_{\frac12}^4\bigr)  \bar O_8  + \bigl( \chi_0^2 \chi_{\frac12}^2 -\chi_{\frac12}^2 \chi_{0}^2\bigr) \bar V_8 \CR
&&\hspace{40mm} +  \bigl( \chi_0 \chi_{\frac12} \chi_0 \chi_{\frac12} -\chi_{\frac12}  \chi_0 \chi_{\frac12} \chi_0  \bigr) \bar S_8+  \bigl( \chi_0 \chi_{\frac12}^2 \chi_0 -\chi_{\frac12}  \chi_0^2 \chi_{\frac12}   \bigr) \bar C_8  \Bigr] \CR
%%%%%%%%%%%%%%%%%%%%%%
&& + \frac12 \bigl(   \mathcal{V} \chi_{\frac12} + \mathcal{H} \chi_{0} \bigr)\bar{\mathcal{Q}} \frac{\Lambda_{\sLambda_{2,2}}[^1_0]}{|\eta^4|^2}  \Bigl[\bigl(  \chi_{\frac12} \chi_0^3 + \chi_0 \chi_{\frac12}^3 \bigr) \bar O_8  + \bigl( \chi_0  \chi_{\frac12} \chi_0^2 + \chi_{\frac12} \chi_0 \chi_{\frac12}^2 \bigr)  \bar V_8 \CR
&& \hspace{62mm} + \bigl( \chi_0^2  \chi_{\frac12} \chi_0 + \chi_{\frac12}^2 \chi_0 \chi_{\frac12} \bigr)   \bar S_8 + \bigl( \chi_0^3  \chi_{\frac12}  + \chi_{\frac12}^3 \chi_0  \bigr)   \bar C_8\Bigr]  
\CR
%%%%%%%%%%%%%%%%%%%%%%
&& + \frac12 \bigl(  \mathcal{V} \chi_{\frac12} - \mathcal{H} \chi_{0} \bigr)\bar{\mathcal{Q}} \frac{\Lambda_{\sLambda_{2,2}}[^1_1]}{|\eta^4|^2} \Bigl[\bigl( - \chi_{\frac12} \chi_0^3 + \chi_0 \chi_{\frac12}^3 \bigr) \bar O_8   \CR
&& \hspace{17mm} + \bigl(   \chi_0  \chi_{\frac12} \chi_0^2 -\chi_{\frac12} \chi_0 \chi_{\frac12}^2 \bigr)  \bar V_8 + \bigl( \chi_0^2  \chi_{\frac12} \chi_0 - \chi_{\frac12}^2 \chi_0 \chi_{\frac12} \bigr)   \bar S_8 + \bigl( \chi_0^3  \chi_{\frac12}  - \chi_{\frac12}^3 \chi_0  \bigr)   \bar C_8\Bigr]     \; . \nonumber \eea
The advantage of this description is that the identification of the non-vanishing momenta along $T^4$ is made manifest. The interpretations of $SU(2)_1$ characters as sums over Hilbert spaces of free fermions allows to identify the isomorphism with the Hilbert space of the `standard' world-sheet fermions. 
To exhibit more explicitly the Hilbert spaces, let us introduce the following notation for the $T^2$ partition functions 
\bea \Lambda_{\sLambda_{2,2}}^{\scalebox{0.6}{even,un} } &=& \frac{1}{2}  \bigl( \Lambda_{\sLambda_{2,2}}[^0_0]+ \Lambda_{ \sLambda_{2,2}}[^0_1]\bigr) \; , \qquad \Lambda_{\sLambda_{2,2}}^{\scalebox{0.6}{odd,un} }=  \frac{1}{2}  \bigl( \Lambda_{\sLambda_{2,2}}[^0_0]-\Lambda_{ \sLambda_{2,2}}[^0_1]\bigr) \; , \CR
 \Lambda_{\sLambda_{2,2}}^{\scalebox{0.6}{even,tw} } &=& \frac{1}{2}  \bigl( \Lambda_{\sLambda_{2,2}}[^1_0]+ \Lambda_{ \sLambda_{2,2}}[^1_1]\bigr) \; , \qquad 
\Lambda_{\sLambda_{2,2}}^{\scalebox{0.6}{odd,tw} } = \frac{1}{2}  \bigl( \Lambda_{\sLambda_{2,2}}[^1_0]-\Lambda_{ \sLambda_{2,2}}[^1_1]\bigr) \; , 
 \eea
which sum all momenta and winding numbers on $T^2$ for which the second momentum is either even or odd and the second winding number is either integer or half-integer. Then one has 
%%%%%%%%%%%%%%%%%%%%%%%%%
\bea \hspace{-5mm}\mathcal{Z}^{T^6/\mathds{Z}_2}_{{\rm II}}  &=&\bar{\mathcal{Q}} 
  \Biggl[ \Bigl(  \mathcal{V} \chi_0    \frac{ \Lambda_{\sLambda_{2,2}}^{\scalebox{0.6}{even,un} }  }{|\eta^4|^2}  + \mathcal{H} \chi_{\frac12}     \frac{ \Lambda_{\sLambda_{2,2}}^{\scalebox{0.6}{odd,un} }  }{|\eta^4|^2}   \Bigr)  \Bigl( \chi_0^4   \bar O_8  + \chi_0^2 \chi_{\frac12}^2  \bar V_8+\chi_0 \chi_{\frac12} \chi_0 \chi_{\frac12}  \bar S_8+  \chi_0 \chi_{\frac12}^2 \chi_0  \bar C_8  \Bigr) \label{TorusPartitionZ2}  \CR
%%%%%%%%%%%%%%%%%%%%%%
&&  +     \Bigl(  \mathcal{V} \chi_0    \frac{ \Lambda_{\sLambda_{2,2}}^{\scalebox{0.6}{odd,un} }  }{|\eta^4|^2}  + \mathcal{H} \chi_{\frac12}     \frac{ \Lambda_{\sLambda_{2,2}}^{\scalebox{0.6}{even,un} }  }{|\eta^4|^2}   \Bigr)   \Bigl(   \chi_{\frac12}^4  \bar O_8  +\chi_{\frac12}^2 \chi_{0}^2 \bar V_8 + \chi_{\frac12}  \chi_0 \chi_{\frac12} \chi_0  \bar S_8+\chi_{\frac12}  \chi_0^2 \chi_{\frac12}   \bar C_8  \Bigr) \CR
%%%%%%%%%%%%%%%%%%%%%%%
&& +  \Bigl(  \mathcal{V} \chi_{\frac12}   \frac{ \Lambda_{\sLambda_{2,2}}^{\scalebox{0.6}{even,tw} }  }{|\eta^4|^2} +  \mathcal{H} \chi_{0}    \frac{ \Lambda_{\sLambda_{2,2}}^{\scalebox{0.6}{odd,tw} }  }{|\eta^4|^2}   \Bigr)  \Bigl( \chi_0 \chi_{\frac12}^3  \bar O_8  + \chi_0  \chi_{\frac12} \chi_0^2   \bar V_8 + \chi_0^2  \chi_{\frac12} \chi_0  \bar S_8 + \chi_0^3  \chi_{\frac12}   \bar C_8\Bigr)  
\\
%%%%%%%%%%%%%%%%%%%%%%
&& +   \Bigl(  \mathcal{V} \chi_{\frac12}   \frac{ \Lambda_{\sLambda_{2,2}}^{\scalebox{0.6}{odd,tw} }  }{|\eta^4|^2} +  \mathcal{H} \chi_{0}    \frac{ \Lambda_{\sLambda_{2,2}}^{\scalebox{0.6}{even,tw} }  }{|\eta^4|^2}   \Bigr)   \Bigl(  \chi_{\frac12} \chi_0^3  \bar O_8  +  \chi_{\frac12} \chi_0 \chi_{\frac12}^2  \bar V_8   + \chi_{\frac12}^2 \chi_0 \chi_{\frac12}    \bar S_8 +  \chi_{\frac12}^3 \chi_0  \bar C_8\Bigr)    \Biggr]   \; . \nonumber \eea
%%%%%%%%%%%%%%%%%%%%%%
Note that one can easily restore the dependence on generic Narain moduli, by identifying the $O(2,6)$ Narain partition functions as
\bea \frac{ \Lambda_{\sLambda_{2,2}}^{\scalebox{0.6}{parity,sec} }  }{|\eta^2|^2} \bar O_8 &\rightarrow& \frac{ \Lambda_{\sLambda_{2,2}\oplus D_4 }^{\scalebox{0.6}{parity,sec} }  }{\eta^2 \bar \eta^6 } \; , \qquad \hspace{4.3mm}\frac{ \Lambda_{\sLambda_{2,2}}^{\scalebox{0.6}{parity,sec} }  }{|\eta^2|^2} \bar V_8 \rightarrow \frac{ \Lambda_{\sLambda_{2,2}\oplus D_4+v }^{\scalebox{0.6}{parity,sec} }  }{\eta^2 \bar \eta^6 } \; , \CR
 \frac{ \Lambda_{\sLambda_{2,2}}^{\scalebox{0.6}{parity,sec} }  }{|\eta^2|^2} \bar S_8 &\rightarrow& \frac{ \Lambda_{\sLambda_{2,2}\oplus D_4+s }^{\scalebox{0.6}{parity,sec} }  }{\eta^2 \bar \eta^6 } \; , \qquad \frac{ \Lambda_{\sLambda_{2,2}}^{\scalebox{0.6}{parity,sec} }  }{|\eta^2|^2} \bar C_8 \rightarrow \frac{ \Lambda_{\sLambda_{2,2}\oplus D_4 +c}^{\scalebox{0.6}{parity,sec} }  }{\eta^2 \bar \eta^6 }  \; .  \eea
The T-duality group of the theory was identified in \cite{Bianchi:2022tbr} as the automorphism group of the lattice $\sLambda_{1,1}\oplus \sLambda_{1,1}[2]\oplus D_4$. It includes the triality automorphism on $D_4^*$ and a Fricke duality that exchanges the winding and momenta $m_2^\prime = 2n_2,n_2^\prime = \frac{m_2}{2}$. We read from the partition function that T-duality permutes states in different affine $SU(2)_1$ modules. Let us write $ \chi_{j_\psi}^\psi , \chi_{j_o}^o ,  \chi_{j_v}^v   , \chi_{j_s}^s ,   \chi_{j_c}^c$ to distinguish the five $SU(2)_1$ characters. We find that in order to act consistently, the Fricke duality must permute $ \chi_{j_\psi}^\psi $ and $ \chi_{j_o}^o $, and  triality  permutes $\chi_{j_v}^v   , \chi_{j_s}^s ,   \chi_{j_c}^c$.\footnote{One does not see directly in the partition function that Fricke duality must permute $ \chi_{j}^\psi $ and $ \chi_{j}^o $ in the $\mathds{Z}_2$ invariant sector with $m_2$ even and $n_2$ integer. Nevertheless, it follows by requiring Fricke invariance of three-point functions. Fricke duality relates the interaction between two massive states with $m_2$ odd and a state of the invariant sector $m_2$ even to the interaction between two massive twisted states and a state in the invariant sector with $ \chi_{j}^\psi $ and $ \chi_{j}^o $ quantum numbers exchanged.}  We show in Appendix \ref{SuperCurrent} that Fricke duality is consistent with $\mathcal{N}=1$ worldsheet supersymmetry. 

Following \cite{Bianchi:2022tbr}, we denote by $\Cong{D}{4}_{0*}(\alpha)$ the full T-duality group, and by $\Cong{D}{4}_{0}(\alpha)$ its subgroup preserving the twisted and the untwisted sectors. $\Cong{D}{4}_{0}(\alpha)$ is a congruent subgroup of the symplectic group $Sp(4,\mathds{H})$ over the Hurwitz quaternions. It is the maximal subgroup of $O(6,6,\mathds{Z})$ preserved by the $\mathds{Z}_2$ orbifold. On the contrary, the Fricke duality arises as an accidental symmetry that mixes the twisted and the untwisted sectors. It will play an important role in the following because it maps geometric branes  to non-geometric branes. 

%%%%%%%%%%%%%%%%%%%%%%%% %%%%%%%%%%%%%%%%%%%%%%%%
\section{Geometric and non-geometric fundamental branes}
Geometric branes can be defined as the orbifold projection of bound states of a brane and its image under the $\mathds{Z}_2$ orbifold action in the original type II theory. A simple example in  type IIA is to consider a D0-brane at a point $(x^4,x^5)$ in $T^2$ and its D4-brane image wrapping $T^4$ at the translated point $(x^4,x^5 {+} \pi R_5)$ in $T^2$.

For simplicity we assume that the Kalb-Ramond field $T_1=0$ on the torus $T^2$, such that the torus partition function factorises as
\be \Lambda_{\sLambda_{2,2}}[^s_r] =  P_{2{[r]}} W_2^{[s]} \; , \ee
with 
\bea P_{2[r]} &=& \sum_{m_1,m_2\in \mathds{Z}} (-1)^{r m_2} e^{ \frac{i \pi \tau }{2 T_2 U_2} | U m_1 -m_2 |^2 -  \frac{i \pi \bar \tau }{2 T_2 U_2} | U m_1 -m_2 |^2} = P^{\scalebox{0.6}{even}}_{2} + (-1)^r P^{\scalebox{0.6}{odd}}_{2} \; , \CR
W_2^{[s]} &=&\sum_{\substack{n_1\in \mathds{Z}\\ n_2 \in \mathds{Z}+\frac{s}{2}}}  e^{ \frac{i \pi \tau  T_2}{2 U_2} | n_1+U n_2 |^2 -  \frac{i \pi \bar \tau T_2 }{2 U_2} | n_1+U n_2|^2}\; . 
\eea
Then the transverse channel amplitude (corresponding to closed-string `tree-level' exchange between two identical D0-D4 branes) only involves momenta along $T^2$ while the winding number of the closed strings are set to zero. It follows that the D0-D4 brane must be geometric, because it cannot couple to the twisted states that all have a non-vanishing half-integer winding number $n_2$. For states with a non-vanishing momentum along $T^4$, one identifies the left momentum with the right momentum $p_L = \pm p_R$. For  vanishing momentum along $T^4$, the states in $ \mathcal{V} \chi_0  P^{\scalebox{0.6}{even}}_{2}  + \mathcal{H} \chi_{\frac12}  P^{\scalebox{0.6}{odd}}_{2}  $ have an even boson mode number in $T^4$, while states in $ \mathcal{V} \chi_0  P^{\scalebox{0.6}{odd}}_{2}  + \mathcal{H} \chi_{\frac12}  P^{\scalebox{0.6}{even}}_{2}  $ have an odd mode number. This gives the transverse channel annulus amplitude  
\be    \widetilde{\mathcal{A}}_{\scalebox{0.7}{D0-D4}} = \frac12 (   \mathcal{V} \chi_0 + \mathcal{H} \chi_{\frac12}  )  \frac{P_2^{[0]} }{T_2 \eta^4} \bigl( O_8  +V_8+S_8 +  C_8   \bigr) + \frac12(   \mathcal{V} \chi_0 - \mathcal{H} \chi_{\frac12}  ) \frac{P_2^{[1]}}{T_2 \eta^4}   \frac{4\eta^2 }{\vartheta_2(0)^2} \; . \ee
To discuss the generalisation to non-geometric branes, it will be useful to write this amplitude using $SU(2)_1$ characters as 
\begin{multline}     \widetilde{\mathcal{A}}_{\scalebox{0.7}{D0-D4}} =  \biggl(  \mathcal{V} \chi_0 \frac{ P^{\scalebox{0.6}{even}}_{2}}{T_2 \eta^4}+\mathcal{H} \chi_{\frac12} \frac{P^{\scalebox{0.6}{odd}}_{2}}{T_2  \eta^4}    \biggr) \Bigl( \chi_0^4 +\chi_0^2 \chi_{\frac12}^2 +  \chi_0\chi_{\frac12} \chi_0 \chi_{\frac12} + \chi_0 \chi_{\frac12}^2 \chi_{0} \Bigr) \\
 +\biggl( \mathcal{H} \chi_{\frac12} \frac{P^{\scalebox{0.6}{even}}_{2}}{T_2  \eta^4}+ \mathcal{V} \chi_0 \frac{P^{\scalebox{0.6}{odd}}_{2}}{T_2  \eta^4} \biggr)  \Bigl(   \chi_{\frac12}^4  + \chi_{\frac12}^2 \chi_0^2 + \chi_{\frac12} \chi_0\chi_{\frac12}\chi_0   +\chi_{\frac12} \chi_{0}^2 \chi_{\frac12} \Bigr) 
 \; . \end{multline}
These characters transform under modular inversion $\tau\rightarrow - 1/\tau$ as 
 \bea  \mathcal{V} \chi_0 \frac{ P^{\scalebox{0.6}{even}}_{2}}{T_2 \eta^4}+\mathcal{H} \chi_{\frac12} \frac{ P^{\scalebox{0.6}{odd}}_{2}}{T_2 \eta^4}  &\rightarrow&  \frac12 \bigl( \mathcal{V} \chi_0 +\mathcal{H} \chi_{\frac12}\bigr)     \frac{ W_{2}^{[0]}}{\eta^4}  +\frac12 \bigl( \mathcal{V} \chi_{\frac12} +\mathcal{H} \chi_{0}\bigr)     \frac{W_{2}^{[1]}}{\eta^4}\; ,  \CR
 \mathcal{H} \chi_{\frac12}\frac{ P^{\scalebox{0.6}{even}}_{2}}{T_2 \eta^4}+  \mathcal{V} \chi_0 \frac{ P^{\scalebox{0.6}{odd}}_{2}}{T_2 \eta^4}  &\rightarrow& \frac12 \bigl( \mathcal{V} \chi_0 +\mathcal{H} \chi_{\frac12}\bigr)     \frac{ W_{2}^{[0]}}{\eta^4}  - \frac12 \bigl( \mathcal{V} \chi_{\frac12} +\mathcal{H} \chi_{0}\bigr)     \frac{ W_{2}^{[1]}}{\eta^4} \; ,  \eea
 and 
\bea  \chi_0^4 +\chi_0^2 \chi_{\frac12}^2 +  \chi_0\chi_{\frac12} \chi_0 \chi_{\frac12} + \chi_0 \chi_{\frac12}^2 \chi_{0}  &\rightarrow&  ( \chi_0 + \chi_{\frac12} ) ( \chi_0^3 +   \chi_{\frac12}^3)  \; , \CR
 \chi_{\frac12}^4  + \chi_{\frac12}^2 \chi_0^2 + \chi_{\frac12} \chi_0\chi_{\frac12}\chi_0   +\chi_{\frac12} \chi_{0}^2 \chi_{\frac12} &\rightarrow& ( \chi_0 - \chi_{\frac12} ) ( \chi_0^3 -   \chi_{\frac12}^3)  \; , \label{Chi4Transform} \eea
 such that the direct channel annulus amplitude is 
 \be  {\mathcal{A}}_{\scalebox{0.7}{D0-D4}} =  \bigl(  \mathcal{V} \chi_0 +\mathcal{H} \chi_{\frac12}\bigr)   \frac{ W_{2}^{[0]}}{\eta^4}   \bigl( \chi_0^4 +\chi_{\frac12}^4 \bigr)  +   \bigl(  \mathcal{V} \chi_{\frac12} +\mathcal{H} \chi_{0}\bigr)   \frac{ W_{2}^{[1]} }{\eta^4}  \bigl(  \chi_{\frac12} \chi_0^3 +  \chi_0 \chi_{\frac12}^3   \bigr)  \; .  \ee 
 The normalisation of the transverse channel amplitude is justified by the interpretation of the direct channel amplitude as a partition function for the open string states. 
 The only contribution to the tension square arises from the NS sector of $\mathcal{V}$ and reads \cite{Harvey:1999gq}
  \be \widetilde{\mathcal{A}}^{{\rm NS}}_{\scalebox{0.7}{D0-D4}}|_{q=0} =    \mathcal{V}^{\rm NS} \chi_0 \frac{P_{2}}{T_2\eta^4} \chi_0^4 \Big|_{q=0}= \frac{4 }{T_2} \; , \ee
  so one gets  the mass $M = \frac{e^{-\phi}}{\sqrt{\alpha^\prime}} $ as expected for a geometric brane. The massless content of the theory is just one $\mathcal{N}=4$ Maxwell multiplet, as expected for a fundamental brane, despite the fact that it only preserves 1/3 of the supersymmetry in $\mathcal{N}=6$ supergravity. 

Although this microscopic description is only valid at a specific locus in moduli, the fundamental D0-D4 brane exists at any point in  Narain moduli space $\Cong{D}{4}_{0}(\alpha)\backslash O^*(8)/U(4)$.

The T-duality group $\Cong{D}{4}_{0*}(\alpha)$ includes the group of unit Hurwitz quaternions $\mathds{H}^\times \subset SU(2)$. One can obtain all the fundamental branes of type D0-D2-D4 by changing the boundary conditions such that the group element $u \in \mathds{H}^\times$ acts on the right sector. In particular $u$ acts on the R-symmetry $SU(2)$ of the $\mathcal{N}=2$ characters $\bar{\mathcal{V}}$ and $\bar{\mathcal{H}}$, and on the right-momentum in $D_4^*$. The open string amplitudes are the all same for identical branes. The generic 1/3 BPS bound states of D0-D2-D4 are more complicated, we will discuss an example in the next section. 

Another example of geometric brane is a D2-brane wrapping $T^2$ and its $\mathds{Z}_2$ image D6-brane wrapping $T^2\times T^4$. Because of the orbifold projection, one would expect nonetheless to have non-geometric branes for which half of $T^2$ is filled with the D2-brane and the other half by the D6-brane. The transverse channel amplitude for this non-geometric brane then couples to the twisted sector since closed string states with fractional winding can couple a D2-brane wrapping half a $T^2$  to a D6-brane wrapping the other half.

By Fricke T-duality, one obtains that the corresponding boundary state is obtained by exchanging the states in $\chi_{j_\psi}^\psi$ and $\chi_{j_o}^o$, in a similar way as for the permutation branes considered in \cite{Gaberdiel:2002jr}. Recall that the action of Fricke T-duality exchanges $\chi_{j_\psi}^\psi$ and $\chi_{j_o}^o$ and momentum and winding $n_2 \rightarrow \frac{m_2}{2},m_2 \rightarrow  2n_2 $, which can be seen to be consistent with  \eqref{TorusPartitionZ2}. This is consistent with worldsheet supersymmetry as we show in Appendix \ref{SuperCurrent}. This gives the transverse channel amplitude 
\begin{multline}     \widetilde{\mathcal{A}}_{\scalebox{0.7}{D2-D6}} = \frac12 \biggl(  \mathcal{V} \chi_0 \frac{T_2 W_{2}^{[0]}}{\eta^4}+\mathcal{H} \chi_{\frac12} \frac{T_2 W_{2}^{[1]}}{\eta^4}    \biggr) \Bigl( \chi_0^4 +\chi_0^2 \chi_{\frac12}^2 +  \chi_0\chi_{\frac12} \chi_0 \chi_{\frac12} + \chi_0 \chi_{\frac12}^2 \chi_{0} \Bigr) \\
 +\frac12 \biggl( \mathcal{H} \chi_{\frac12} \frac{T_2 W_{2}^{[0]} }{\eta^4}+ \mathcal{V} \chi_0 \frac{T_2 W_{2}^{[1]}}{\eta^4} \biggr)  \Bigl(  \chi_{\frac12}^4  + \chi_{\frac12}^2 \chi_0^2 + \chi_{\frac12} \chi_0\chi_{\frac12}\chi_0   +\chi_{\frac12} \chi_{0}^2 \chi_{\frac12}\Bigr) 
 \; . \end{multline}
 The corresponding Ishibashi state is similar to the one constructed in \cite{Brunner:1999fj} in the trivial orbifold theory, so that $\mathcal{H} \chi_\frac12 \chi_0^4 $  is identified on the left-moving sector to $\mathcal{H} \chi^o_{\frac12}\chi_0^\psi \chi^v_0\chi^s_0\chi^c_0 $ where the two  vacua of the bosonic module $\chi^o_{\frac12}\chi^v_0\chi^s_0\chi^c_0 $ are generated by the  $SU(2)$ doublet of twisted boson fields. 

 The modular inversion of $\tau$ gives 
 \bea  \mathcal{V} \chi_0 \frac{T_2 W_{2}^{[0]}}{\eta^4}+\mathcal{H} \chi_{\frac12} \frac{T_2 W_{2}^{[1]}}{\eta^4}  &\rightarrow&  \bigl( \mathcal{V} \chi_0 +\mathcal{H} \chi_{\frac12}\bigr)     \frac{ P^{\scalebox{0.6}{even}}_{2}}{\eta^4}  + \bigl( \mathcal{V} \chi_{\frac12} +\mathcal{H} \chi_{0}\bigr)     \frac{ P^{\scalebox{0.6}{odd}}_{2}}{\eta^4}\; ,  \CR
 \mathcal{H} \chi_{\frac12}\frac{T_2 W_{2}^{[0]}}{\eta^4}+  \mathcal{V} \chi_0 \frac{T_2 W_{2}^{[1]}}{\eta^4}  &\rightarrow&  \bigl( \mathcal{V} \chi_0 +\mathcal{H} \chi_{\frac12}\bigr)     \frac{ P^{\scalebox{0.6}{even}}_{2}}{\eta^4}  - \bigl( \mathcal{V} \chi_{\frac12} +\mathcal{H} \chi_{0}\bigr)     \frac{ P^{\scalebox{0.6}{odd}}_{2}}{\eta^4} \; ,  \eea
 and \eqref{Chi4Transform} such that the direct channel amplitude is 
 \be  {\mathcal{A}}_{\scalebox{0.7}{D2-D6}} =  \bigl(  \mathcal{V} \chi_0 +\mathcal{H} \chi_{\frac12}\bigr)   \frac{ P^{\scalebox{0.6}{even}}_{2}}{\eta^4}   \bigl( \chi_0^4 +\chi_{\frac12}^4 \bigr)  +   \bigl(  \mathcal{V} \chi_{\frac12} +\mathcal{H} \chi_{0}\bigr)   \frac{ P^{\scalebox{0.6}{odd}}_{2}}{\eta^4}  \bigl(  \chi_{\frac12} \chi_0^3 +  \chi_0 \chi_{\frac12}^3   \bigr)  \; .  \ee 
 The only contribution to the tension square arises once again from the NS sector of $\mathcal{V}$ and reads 
  \be \widetilde{\mathcal{A}}^{{\rm NS}}_{\scalebox{0.7}{D2-D6}}|_{q=0} = \frac12   \mathcal{V}^{\rm NS} \chi_0 \frac{T_2 W_{2}^{[0]}}{\eta^4} \chi_0^4 \Big|_{q=0}= 2 T_2 \; , \ee
  so one gets the mass $M = \frac{e^{-\phi} T_2}{\sqrt{2 \alpha^\prime}}$, which is $\frac{1}{\sqrt{2}}$ the expected mass for a geometric D2-D6 brane. The massless content of the theory is just one $\mathcal{N}=4$ Maxwell multiplet without additional hyper-multiplets, as for the geometric D0-D4 brane. The factor of $\frac{1}{\sqrt{2}}$ in the tension is the same as the one found in \cite{Gaberdiel:2002jr} for a similar asymmetric orbifold of the bosonic string. It is consistent with the Euclidean D-brane instantons corrections obtained in \cite{Bianchi:2022tbr}, based on supersymmetry and U-duality. The factor of $\frac{1}{\sqrt{2}}$ follows by Fricke T-duality, because the general D0-D4 charge $q\in \mathds{H}$ is mapped under Fricke T-duality to the D2-D6 charge $\frac{1}{\alpha} q$ with $|\alpha|^2 = 2$. Although this microscopic description is only valid at a specific locus in moduli space, it is obtained by Fricke T-duality from a geometric brane and we therefore expect this fundamental D2-D6 brane to exist at any point in  Narain moduli space.

\section{Geometric D2-D6 brane  as a bound state}
Let us now consider a D0-D2-D4 brane. The direct construction of the boundary state leads to a supersymmery breaking brane with a scalar tachyon. Just as for the D0-D2 boundary state in type II on a torus, one expects the transverse channel amplitude between D0-D4 and D2-D2 to give a phase associated to the $SO(2)_1$ character such that \footnote{For a D0-D2 brane one would use $V_8 = V_6 O_2+O_6 V_2$ and attribute a minus sign to $V_2$.}
\be \mathcal{V} \chi_0 \rightarrow \bigl( V_4 \chi_0(\tfrac12) - S_4 \chi_{\frac12}(\tfrac12) \bigr) \chi_0(\tfrac12) \;, \qquad  \mathcal{H} \chi_{\frac12} \rightarrow \bigl( O_4 \chi_{\frac12}(\tfrac12) - C_4 \chi_{0}(\tfrac12) \bigr) \chi_{\frac12}(\tfrac12) \;. \ee
For short we define 
\be \mathcal{V}(z) = V_4 \chi_0(z) - S_4 \chi_{\frac12}(z) \; , \qquad \mathcal{H}(z) = O_4 \chi_{\frac12}(z) - C_4 \chi_{0}(z)\; . \ee
The contribution of $\mathcal{V}(\frac12) $ to the tension is the same as for $\mathcal{V}$, while its contribution to the Ramond-Ramond (RR) charge vanishes. It results an attractive force that should lead to the recombination of the branes into a supersymmetric D0-D2-D4 bound state.

The argument $z= \frac{1}{2}$  of the $SU(2)_1$ character in $\mathcal{V}$ and $\mathcal{H}$ is understood as an order four  element $i \sigma_3\in SU(2)$.  The RR charges of the D0-D4 and the D2-D2 branes are related by this order four $SU(2)$ element interpreted as a pure imaginary unit in $\mathds{H}^\times$.  As a T-duality, it acts on the $T^4$ sector by exchanging two pairs of $SU(2)_1$ affine modules, that we choose to be $\chi^o_j\leftrightarrow \chi^v_j $ and $\chi^s_j  \leftrightarrow \chi^c_j$. It follows that the transverse amplitude between  D0-D4 and D2-D2 can only involve non-zero momenta in $D_4$ and $D_4+v$ (of respective affine characters  $O_8$ and $V_8$). The transverse channel amplitude is 
\begin{multline}    \widetilde{\mathcal{A}}_{\scalebox{0.7}{D0-D4}\rightarrow \scalebox{0.7}{D2-D2}} = \frac12 \bigl(   \mathcal{V}(\tfrac12)  \chi_0(\tfrac12) + \mathcal{H}(\tfrac12) \chi_{\frac12}(\tfrac12)  \bigr)  \frac{P_2^{[0]} }{T_2 \eta^4} \bigl( O_2+V_2   \bigr)^3 \bigl( O_2-V_2   \bigr) \\ + \frac12\bigl(   \mathcal{V}(\tfrac12) \chi_0(\tfrac12) - \mathcal{H}(\tfrac12) \chi_{\frac12}(\tfrac12)  \bigr) \frac{P_2^{[1]}}{T_2 \eta^4}   \frac{4\eta^2 }{\vartheta_2(0)^2} \; , \end{multline}
where 
\be (O_2+V_2  )^3 (O_2-V_2 ) = \frac{\sum_{p\in D_4} e^{i \pi \tau (p,p) + \pi i (\mu,p)} +\sum_{p\in D_4+v} e^{i \pi \tau (p,p) + \pi i (\mu,p)} }{\eta^4} \ee
for $\mu \in D_4+v$ of norm square $(\mu,\mu) = 1$, is interpreted as a Wilson line insertion in $O_8 + V_8$.\footnote{The same Wilson line gives zero if inserted in $S_8$ and $C_8$, and so consistently projects out the states that are incompatible with the mixed Dirichlet--Neumann boundary conditions.} In the representation of the $SU(2)_1$ modules in terms of free fermions, the T-duality permutes the $SU(2)_1$ characters $\chi^o_j\leftrightarrow \chi^v_j $ and $\chi^s_j  \leftrightarrow \chi^c_j$ and gives a transverse amplitude associated to a permutation brane \cite{Recknagel:2002qq}
\begin{multline}     \widetilde{\mathcal{A}}_{\scalebox{0.7}{D0-D4}\rightarrow \scalebox{0.7}{D2-D2}} =  \biggl(  \mathcal{V}(\tfrac12) \chi_0(\tfrac12) \frac{ P^{\scalebox{0.6}{even}}_{2}}{T_2 \eta^4}+\mathcal{H}(\tfrac12) \chi_{\frac12}(\tfrac12) \frac{P^{\scalebox{0.6}{odd}}_{2}}{T_2  \eta^4}    \biggr) \chi_0(2\tau)  \bigl( \chi_0(2\tau) + \chi_{\frac12}(2\tau)\bigr)   \\
 +\biggl( \mathcal{H}(\tfrac12) \chi_{\frac12}(\tfrac12) \frac{P^{\scalebox{0.6}{even}}_{2}}{T_2  \eta^4}+ \mathcal{V}(\tfrac12) \chi_0(\tfrac12) \frac{P^{\scalebox{0.6}{odd}}_{2}}{T_2  \eta^4} \biggr)   \chi_{\frac12}(2\tau)  \bigl( \chi_0(2\tau) + \chi_{\frac12}(2\tau)\bigr)  \; , \end{multline}
 where $ \chi_j(2\tau)$ appears because the two states in the two $SU(2)_1$ modules must be the same. 
One checks indeed that \footnote{These identities follow from $\frac{\vartheta_4(2\tau)}{\eta(2\tau)} = \frac{2\eta(\tau)}{\vartheta_2(\tau)}$, $\frac{\vartheta_3(2\tau)^2}{\eta(2\tau)^2} = \frac{\vartheta_3(\tau)^3 \vartheta_4(\tau) +  \vartheta_3(\tau) \vartheta_4(\tau)^3}{2\eta(\tau)^4}$ and $\frac{\vartheta_2(2\tau)^2}{\eta(2\tau)^2} = \frac{\vartheta_3(\tau)^3 \vartheta_4(\tau) -  \vartheta_3(\tau) \vartheta_4(\tau)^3}{2\eta(\tau)^4}$ .}
\be \bigl( O_2+V_2   \bigr)^3 \bigl( O_2-V_2   \bigr)  =  \bigl( \chi_0(2\tau) + \chi_{\frac12}(2\tau)\bigr) ^2\; , \qquad  \bigl( \chi_0(2\tau)^2 - \chi_{\frac12}(2\tau)^2\bigr) =   \frac{4\eta^2 }{\vartheta_2(0)^2}\; , \ee
so the two interpretations are compatible. 

The direct channel amplitude is 
\begin{multline}     {\mathcal{A}}_{\scalebox{0.7}{D0-D4}\rightarrow \scalebox{0.7}{D2-D2}} =  \Bigl[ ( V_4 \chi_{\frac14} - S_4 \chi_{-\frac14} ) \chi_{\frac14}  +( O_4 \chi_{-\frac14} - C_4 \chi_{\frac14} ) \chi_{-\frac14} \Bigr]   \frac{ W^{[0]}_{2}}{ \eta^4}   \chi_{0}(\tfrac{\tau}{2})\chi_0(\tfrac{\tau}{2})   \\
 +\Bigl[ ( V_4 \chi_{\frac14} - S_4 \chi_{-\frac14} ) \chi_{-\frac14}  +( O_4 \chi_{-\frac14} - C_4 \chi_{\frac14} ) \chi_{\frac14} \Bigr]   \frac{ W^{[1]}_{2}}{ \eta^4}    \chi_{\frac12}(\tfrac{\tau}{2})\chi_0(\tfrac{\tau}{2}) \; , \end{multline}
which is consistent with the interpretation as a Hilbert space trace, and where we use the notation \eqref{SU2Char} for $j= \pm \frac14$, in which case $\chi_{\pm \frac14} = \xi_{\pm 1}^4 + \xi_{\mp 3}^4$ is not an $SU(2)_1$ character but a short for the sum of two $U(1)_1$ characters \eqref{SU2U1}. One finds one scalar tachyon coming from $O_4 \chi_{-\frac14}^2  $ that is expected to drive the system to a supersymmetric bound state. 

We will determine this bound state at the end of this section, but let us first describe the amplitude between two non-geometric branes.

Using Fricke T-duality one can directly obtain the dual amplitude between a non-geometric D2-D6 and a non-geometric D4-D4 brane wrapping  half $T^2$. The transverse channel amplitude is 
\begin{multline}     \widetilde{\mathcal{A}}_{\scalebox{0.7}{D2-D6}\rightarrow \scalebox{0.7}{D4-D4}} =  \frac12 \biggl(  \mathcal{V}(\tfrac12) \chi_0(\tfrac12) \frac{ T_2 W^{[0]}_{2}}{ \eta^4}+\mathcal{H}(\tfrac12) \chi_{\frac12}(\tfrac12) \frac{T_2 W^{[1]}_{2}}{  \eta^4}    \biggr) \chi_0(2\tau)  \bigl( \chi_0(2\tau) + \chi_{\frac12}(2\tau)\bigr)   \\
 +\frac12 \biggl( \mathcal{H}(\tfrac12) \chi_{\frac12}(\tfrac12) \frac{T_2 W^{[0]}_{2} }{  \eta^4}+ \mathcal{V}(\tfrac12) \chi_0(\tfrac12) \frac{T_2 W^{[1]}_{2} }{ \eta^4} \biggr)   \chi_{\frac12}(2\tau)  \bigl( \chi_0(2\tau) + \chi_{\frac12}(2\tau)\bigr)  \; , \end{multline}
 and the direct channel amplitude 
 \begin{multline}     {\mathcal{A}}_{\scalebox{0.7}{D2-D6}\rightarrow \scalebox{0.7}{D4-D4}} =  \Bigl[ ( V_4 \chi_{\frac14} - S_4 \chi_{-\frac14} ) \chi_{\frac14}  +( O_4 \chi_{-\frac14} - C_4 \chi_{\frac14} ) \chi_{-\frac14} \Bigr]   \frac{ P^{\scalebox{0.6}{even}}_{2}}{ \eta^4}   \chi_{0}(\tfrac{\tau}{2})\chi_0(\tfrac{\tau}{2})   \\
 +\Bigl[ ( V_4 \chi_{\frac14} - S_4 \chi_{-\frac14} ) \chi_{-\frac14}  +( O_4 \chi_{-\frac14} - C_4 \chi_{\frac14} ) \chi_{\frac14} \Bigr]   \frac{ P^{\scalebox{0.6}{odd}}_{2} }{ \eta^4}    \chi_{\frac12}(\tfrac{\tau}{2})\chi_0(\tfrac{\tau}{2}) \; . \end{multline}
Perturbative techniques do not allow directly to determine the supersymmetric boundary state that should emerge from tachyon condensation. However, the unique supersymmetric object with the correct tension square 
  \be \widetilde{\mathcal{A}}^{{\rm NS}}_{\scalebox{0.7}{D2-D4-D6}}|_{q=0} =    \mathcal{V}^{\rm NS} \chi_0 \frac{T_2 W_{2}^{[0]}}{\eta^4} \chi_0^4 \Big|_{q=0}= 4 T_2 \; , \ee
and RR charge is a geometric BPS brane. Repeating the construction of the previous section one obtains the transverse channel amplitude of the geometric D2-D4-D6 brane 
\begin{multline}     \widetilde{\mathcal{A}}_{\scalebox{0.7}{D2-D4-D6}} =  \mathcal{V} \chi_0 \frac{T_2 W_{2}^{[0]}}{\eta^4} \Bigl( \chi_0^4 +\chi_0^2 \chi_{\frac12}^2 +  \chi_0\chi_{\frac12} \chi_0 \chi_{\frac12} + \chi_0 \chi_{\frac12}^2 \chi_{0} \Bigr) \\
 + \mathcal{H} \chi_{\frac12} \frac{T_2 W_{2}^{[0]} }{\eta^4}  \Bigl(  \chi_{\frac12}^4  + \chi_{\frac12}^2 \chi_0^2 + \chi_{\frac12} \chi_0\chi_{\frac12}\chi_0   +\chi_{\frac12} \chi_{0}^2 \chi_{\frac12}\Bigr) 
 \; ,\end{multline}
and its direct channel amplitude 
\be   {\mathcal{A}}_{\scalebox{0.7}{D2-D4-D6}}  =  \bigl(  \mathcal{V} \chi_0 +\mathcal{H} \chi_{\frac12}\bigr)   \frac{ P_{2}}{\eta^4}   \bigl( \chi_0^4 +\chi_{\frac12}^4 \bigr)  +   \bigl(  \mathcal{V} \chi_{\frac12} +\mathcal{H} \chi_{0}\bigr)   \frac{ P_{2}}{\eta^4}  \bigl(  \chi_{\frac12} \chi_0^3 +  \chi_0 \chi_{\frac12}^3   \bigr)  \; .  \ee
In this case there is no distinction between odd and even momenta $m_2$, and the massless content of the theory includes both an $\mathcal{N}=4$ Maxwell multiplet and an $\mathcal{N}=2$ hyper-multiplet.\footnote{The choice of a fundamental D2-D6 brane being non-geometric and the geometric bound state a D2-D4-D6 is a priori conventional. But they cannot be both D0-D4 as they carry different RR charges.}

By Fricke T-duality one can now deduce the supersymmetric D0-D2-D4 bound state transverse channel amplitude as being
\begin{multline}     \widetilde{\mathcal{A}}_{\scalebox{0.7}{D0-D2-D4}} = 2 \mathcal{V} \chi_0 \frac{P^{\scalebox{0.6}{even}}_{2}}{T_2 \eta^4} \Bigl( \chi_0^4 +\chi_0^2 \chi_{\frac12}^2 +  \chi_0\chi_{\frac12} \chi_0 \chi_{\frac12} + \chi_0 \chi_{\frac12}^2 \chi_{0} \Bigr) \\
 +2 \mathcal{H} \chi_{\frac12} \frac{ P^{\scalebox{0.6}{even}}_{2}}{T_2 \eta^4}  \Bigl(  \chi_{\frac12}^4  + \chi_{\frac12}^2 \chi_0^2 + \chi_{\frac12} \chi_0\chi_{\frac12}\chi_0   +\chi_{\frac12} \chi_{0}^2 \chi_{\frac12}\Bigr) 
 \; ,\end{multline}
and the direct channel amplitude 
\be   {\mathcal{A}}_{\scalebox{0.7}{D0-D2-D4}}  =  \bigl(  \mathcal{V} \chi_0 +\mathcal{H} \chi_{\frac12}\bigr)   \frac{ W_2^{[0]} {+}W_2^{[1]}  }{\eta^4}   \bigl( \chi_0^4 +\chi_{\frac12}^4 \bigr)  +   \bigl(  \mathcal{V} \chi_{\frac12} +\mathcal{H} \chi_{0}\bigr)   \frac{ W_2^{[0]} {+}W_2^{[1]}}{\eta^4}  \bigl(  \chi_{\frac12} \chi_0^3 +  \chi_0 \chi_{\frac12}^3   \bigr)  \; .  \ee
Note that although the transverse amplitude above does not involve the twisted sector, it is non-geometric in the sense that the left-moving fermion and boson modules $\chi_j^\psi$ and $\chi_j^o$ are permuted in the definition of the Ishibashi state.

This provides the two simplest examples of composite BPS branes in the asymmetric orbifold. In this case we could argue from Fricke T-duality what the direct amplitude should be. For more general cases such a short argument will not work. It may nonetheless be possible to determine the composite geometric branes by relying on duality symmetry. One may expect for example that more general D0-D2-D4 brane might be described by a D0-D4 brane in a magnetic field.\footnote{We thank C.~Angelantonj for suggesting this interpretation.} 
 
 \section{The  $\mathds{Z}_3$ asymmetric orbifold}
 One can do a similar computation for the $\mathds{Z}_3$ orbifold at the $A_2\oplus A_2$ symmetric point \eqref{A2A2} \cite{Bianchi:2022tbr}. For simplicity we  consider again the Narain moduli such that 
\be \frac{\Lambda_{\sLambda_{6,6}}}{|\eta|^{12}} = \frac{\Lambda_{\sLambda_{2,2}}}{|\eta|^4}   \biggl(\sum_{p=0}^2 |  \chi^{\scalebox{0.6}{su(3)}}_{p}  |^2 \biggr)^2\; , \ee
where $ \chi^{\scalebox{0.6}{su(3)}}_{p}$ for $p=0,1,2$ are the three $SU(3)_1$ characters respectively associated to the trivial, the fundamental and the anti-fundamental representations of $SU(3)$. 

The torus partition function in the $\mathds{Z}_3$ orbifold can be written 
 \bea \mathcal{Z}^{T^6/\mathds{Z}_3}_{{\rm II}}   &=&\frac13 \bigl(  \mathcal{V} \chi_0 + \mathcal{H} \chi_{\frac12} \bigr) \bar{\mathcal{Q}}   \frac{\Lambda_{\sLambda_{2,2}}}{|\eta^4|^2} \biggl(\sum_{p=0}^2 |  \chi^{\scalebox{0.6}{su(3)}}_{p}  |^2 \biggr)^2  \CR
 &&   + \frac13 \bigl(  \mathcal{V} \chi_0(2/3) - \mathcal{H} \chi_{\frac12}(2/3)  \bigr) \bar{\mathcal{Q}}    \frac{\Lambda_{\sLambda_{2,2}}[\omega^m]}{|\eta^4|^2} \frac{3\eta^2 }{\vartheta_1(\frac13)^2}  \bar{\chi}_0^{\scalebox{0.6}{su(3)}} \, \bar{\chi}_0^{\scalebox{0.6}{su(3)}}   \CR
 &&   + \frac13 \bigl(  \mathcal{V} \chi_0(1/3) - \mathcal{H} \chi_{\frac12}(1/3)  \bigr)  \bar{\mathcal{Q}}   \frac{\Lambda_{\sLambda_{2,2}}[\bar \omega^m]}{|\eta^4|^2}  \frac{3\eta^2 }{\vartheta_1(\frac23)^2}  \bar{\chi}_0^{\scalebox{0.6}{su(3)}} \, \bar{\chi}_0^{\scalebox{0.6}{su(3)}}   \eea
 with $\omega = e^{\frac{2\pi i}{3}}$. It will be convenient to write the partition function in terms of $U(1)_1$ characters in order to exhibit the modules involved and their chiral algebra symmetry. The level 1 $SU(3)$ characters decompose into the products of  level 1 $SU(2)$ characters and  level 1 $U(1)$ characters according to 
\be \chi_0^{\scalebox{0.6}{su(3)}} = \chi_0 \xi^3_0 + \chi_{\frac12} \xi^3_3\; , \quad \chi_1^{\scalebox{0.6}{su(3)}}  = \chi_0 \xi^3_2 + \chi_{\frac12} \xi^3_5\; , \quad \chi_2^{\scalebox{0.6}{su(3)}} = \chi_0 \xi^3_4 + \chi_{\frac12} \xi^3_1\; ,\ee
where each $U(1)_1$ character $\xi^3_p$ corresponds to the $\mathcal{A}_3$ modules of the primary fields $e^{ i \frac{p}{\sqrt{3}} \varphi} $. Recall that for a free field $\varphi$ of radius $R=\sqrt{N}$, i.e. defined modulo $2\pi \sqrt{N}$, one defines the dimension $\frac{p^2}{4N}$ vertex operators \footnote{Notice that the free field is normalised such that $\partial \varphi(z)  \partial \varphi(w) \sim \frac{-1/2}{(z-w)^2}$.} 
\be V_p = e^{ i \frac{p}{\sqrt{N}} \varphi} \; , \ee
that belong to  irreducible modules of the chiral algebra $\mathcal{A}_N$ generated by the current $J = i \partial \varphi$ and the vertex operators $V_{\pm 2N}$, for each $p$ modulo $2N$ \cite{Dijkgraaf:1989hb}.

One also computes that the partition function of the $T^4$ free bosons with insertion of the $\mathds{Z}_3$ generator can be rewritten using 
\be \frac{\sqrt{3}\eta  }{\vartheta_1(\frac13)} =  \chi_0(1/3) \xi^3_0 - \chi_{\frac12}(1/3)  \xi^3_3 \; , \qquad \chi_0(1/3) \xi^3_2 - \chi_{\frac12}(1/3) \xi^3_5 = 0 \; . \ee
The orbifold projection preserves the symmetry algebra $\mathcal{A}_3 \times \mathcal{A}_3$ associated to the the two $U(1)$ currents with radius $R= \sqrt{3}$. For the three $U(1)$ currents with radius $R=3$, the symmetry algebra preserved by the diagonal $\mathds{Z}_3$ action  on $SU(2)_1^3$ is generated by the three $U(1)$ currents and the vertex operators of integer dimension 
\be V^1_{\pm 6} V^2_{\pm 6} V^3_{\pm 6}\; , \quad V^i_{\pm 12} V^j_{\pm 6} \; ,  \quad V^i_{\pm 18}\; , \ee
with $V^i_p = e^{ i \frac{p}{3} \varphi_i}$. We will call this symmetry chiral algebra $\mathcal{A}^3_9$. By definition it includes  $ \mathcal{A}_9 \times \mathcal{A}_9\times \mathcal{A}_9$, and to identify the irreducible modules it will be convenient to decompose the characters $\chi_j(z)$ above into $\xi_p^9$ characters as  
 \be \chi_0 = \xi_{0}^9 + \xi_6^9 + \xi_{12}^9  \; , \quad  \chi_{\frac12} = \xi_{3}^9+\xi_{9}^9+\xi_{15}^9\; ,  \ee
 and 
 \be \chi_0(1/3) = \xi_{0}^9 + \omega \xi_6^9 +\omega^2  \xi_{12}^9 \; , \quad -\chi_{\frac12}(1/3) = \omega^2 \xi_{3}^9 + \xi_9^9 +\omega   \xi_{15}^9 \; . \ee

One can use these formulas together with \eqref{thetaxixi} to write the  $\mathds{Z}_3$ orbifold partition function on the torus 
 \bea \mathcal{Z}^{T^6/\mathds{Z}_3}_{{\rm II}}   &=&\frac13 \sum_{r=0}^2 \Biggl\{ \bigl[  \mathcal{V} ( \xi_0^9 + \omega^r \xi_6^9 + \bar \omega^r \xi_{12}^9 ) + \mathcal{H} ( \xi_9^9 + \omega^r \xi_{15}^9 + \bar \omega^r \xi_{3}^9 )  \bigr] \bar{\mathcal{Q}}  \frac{\Lambda_{\sLambda_{2,2}}[^0_r]}{|\eta^4|^2}   \\
 && \times  \biggl(\sum_{p=0}^2 \Bigl[  ( \xi_0^9 + \omega^r \xi_6^9 + \bar \omega^r \xi_{12}^9 ) \xi_{2p}^3 + ( \xi_9^9 + \omega^r \xi_{15}^9 + \bar \omega^r \xi_{3}^9 )   \xi_{2p+3}^3 \Bigr] \bar{\chi}_p^{\scalebox{0.6}{su(3)}}  \biggr)^2 \Biggr\} \CR 
 && \hspace{-15mm} + \frac13 \sum_{s=1}^2 \sum_{r=0}^2 \Biggl\{ \bigl[  \mathcal{V} ( \xi_{4s}^9 {+} \omega^{r s} \xi_{4s+6}^9 {+} \bar \omega^{r s}  \xi_{4s+12}^9 ) + \mathcal{H} ( \xi_{4s+9}^9 {+} \omega^{r s}  \xi_{4s-3}^9 {+}  \bar\omega^{r s}\xi_{4s+3}^9 )  \bigr] \bar{\mathcal{Q}} \frac{\Lambda_{\sLambda_{2,2}}[^s_r]}{|\eta^4|^2}   \CR
 && \hspace{-7mm} \times\omega^r   \biggl(\sum_{p=0}^2 \Bigl[  ( \xi_{4s}^9 {+} \omega^{r s} \xi_{4s+6}^9 {+} \bar \omega^{r s}  \xi_{4s+12}^9  ) \xi_{2p}^3 + (  \xi_{4s+9}^9{ +} \omega^{r s}  \xi_{4s-3}^9 {+}  \bar\omega^{r s}\xi_{4s+3}^9 )   \xi_{2p+3}^3 \Bigr] \bar{\chi}_p^{\scalebox{0.6}{su(3)}}  \biggr)^2 \Biggr\} \nonumber\; .   \eea
To exhibit more explicitly the modules involved, it is convenient to introduce the notation 
\bea \Lambda_{\sLambda_{2,2}}^{r,s } &=& \frac{1}{3}  \bigl( \Lambda_{\sLambda_{2,2}}[^s_0]+ \bar \omega^{r}  \Lambda_{ \sLambda_{2,2}}[^s_1]+ \omega^{r}  \Lambda_{ \sLambda_{2,2}}[^s_2]\bigr)\CR
&=&  \sum_{\substack{m_1,n_1\in \mathds{Z}\\m_2\in 3 \mathds{Z} + r\\  n_2 \in \mathds{Z}+\frac{s}{3}  }} e^{ \frac{i \pi \tau }{2 T_2 U_2} | U m_1 -m_2 + T (n_1+U n_2)|^2 -  \frac{i \pi \bar \tau }{2 T_2 U_2} | U m_1 -m_2 + \bar T (n_1+U n_2)|^2}  \; , \eea
such that 
 \begin{multline} \mathcal{Z}^{T^6/\mathds{Z}_3}_{{\rm II}}   = \hspace{-5mm} \sum_{\substack{r,s,\ell_1,\ell_2,\ell_3,p,q \in \{ 0,1,2\} \\ \ell_1+\ell_2+\ell_3+r+s= 0\; {\rm mod}\;  3 }}\sum_{k_1,k_2\in\{0,1\}} \bigl( \mathcal{V} \xi^9_{4s+6\ell_1} + \mathcal{H} \xi^9_{4s+6\ell_1+9}\bigr)  \xi^9_{4s+6\ell_2+9k_1}\xi^9_{4s+6\ell_3+9k_2 }  \\
  \times \xi^3_{2p+3k_1}\xi^3_{2q+3k_2}  \frac{ \Lambda_{\sLambda_{2,2}}^{r,s } }{|\eta^4|^2} \bar{\mathcal{Q}} \,  \bar{\chi}_p^{\scalebox{0.6}{su(3)}}  \, \bar{\chi}_q^{\scalebox{0.6}{su(3)}}   \; ,  \label{TorusZ3} \end{multline}
with 
\be  \bar{\mathcal{Q}} \,  \bar{\chi}_p^{\scalebox{0.6}{su(3)}}  \, \bar{\chi}_q^{\scalebox{0.6}{su(3)}}   =  \sum_{\ell_1,\ell_2,\ell_3 \in \{ 0,1,2\}}\sum_{k_1,k_2\in\{0,1\}} \bigl( \bar{\mathcal{V}} \bar{\xi}^9_{6\ell_1} + \bar{\mathcal{H}} \bar{\xi}^9_{6\ell_1+9}\bigr)  \bar{\xi}^9_{6\ell_2+9k_1}\bar{\xi}^9_{6\ell_3+9k_2 } \bar{\xi}^3_{2p+3k_1}\bar{\xi}^3_{2q+3k_2} \; .   \ee
 One identifies the 72 irreducible $\mathcal{A}_9^3$ modules of characters 
 \be \xi_9^3[^{r,s}_{\; k_i}] = \sum_{\substack{\ell_1,\ell_2,\ell_3 \in \{ 0,1,2\} \\ \ell_1+\ell_2+\ell_3+r+s= 0\; {\rm mod}\;  3 }} \xi^9_{4s+6 \ell_1 + 9 k_1} \xi^9_{4s+6 \ell_2 + 9 k_2} \xi^9_{4s+6 \ell_3 + 9 k_3} \; ,  \label{A39Char} \ee 
 labeled by $k_i \in \{0,1\}$ and $r,s\in \{0,1,2\}$.

 \subsection{The geometric D0-D4 brane}

We can now consider the transverse channel amplitude for a D0-D4  boundary state. Let us first address the geometric description of the boundary state. For non-zero momentum along $T^4$, the orbifold projection identifies the momenta in triplets and one simply gets a 1/3 factor in the amplitude. For vanishing momentum along $T^4$, the $T^4$ free boson mode number must have a fixed congruence modulo $3$ and we get 
\begin{multline} \widetilde{\mathcal{A}}_{\scalebox{0.7}{D0-D4}}  = \frac{1}{3} \bigl(   \mathcal{V} \chi_0 + \mathcal{H} \chi_{\frac12}  \bigr) \frac{P_{2}^{[0]}}{T_2 \eta^4}\bigl(  \chi_{[0,0]}  +\chi_{[1,0]}  +\chi_{[0,1]}\bigr)^2 + \frac13 (  \mathcal{V} \chi_0(1/3) - \mathcal{H} \chi_{\frac12}(1/3)  )    \frac{P_{2}^{[1]}}{T_2 \eta^4} \frac{3\eta^2 }{\vartheta_1(\frac13)^2} \\
+   \frac13 (  \mathcal{V} \chi_0(2/3) - \mathcal{H} \chi_{\frac12}(2/3)  )   \frac{P_{2}^{[2]}}{T_2 \eta^4} \frac{3\eta^2 }{\vartheta_1(\frac23)^2} \; . \end{multline}
It will be useful to write this amplitude using $U(1)_1$ characters as 
%One writes the transverse channel amplitude 
\be  \widetilde{\mathcal{A}}_{\scalebox{0.7}{D0-D4}}  = \sum_{r=0}^2  \frac{P_2^{r\, {\rm mod}\, 3}}{T_2\eta^4}  \hspace{-4mm} \sum_{p,p^\prime\in \{ 0,1,2\}}  \hspace{-5mm} \sum_{\substack{ \ell_i \in \{ 0,1,2\} \\ \ell_1+\ell_2+\ell_3+r = 0 \; {\rm mod}\;  3}}  \hspace{-3mm} \sum_{k,k^\prime\in\{0,1\}} \hspace{-2mm} ( \mathcal{V} \xi^9_{6\ell_1} + \mathcal{H} \xi^9_{6\ell_1+9})  \xi^9_{6\ell_2+9k}\xi^9_{6\ell_3+9k^\prime } \xi^3_{2p+3k}\xi^3_{2p^\prime+3k^\prime} \; ,  \ee
where we introduced 
\be P_2^{r\, {\rm mod}\, 3} =\sum_{\substack{m_1\in \mathds{Z}\\ m_2 \in 3\mathds{Z}+r }} e^{ \frac{i \pi \tau }{2 T_2 U_2} | U m_1 -m_2 |^2 -  \frac{i \pi \bar \tau }{2 T_2 U_2} | U m_1 -m_2 |^2} \ee
such that 
\be P_{2[r]} = P_2^{0\, {\rm mod}\, 3} + \omega^r P_2^{1\, {\rm mod}\, 3} +\bar \omega^r P_2^{2\, {\rm mod}\, 3}  \; . \ee
Using a modular inversion one gets the direct channel amplitude 
\bea  {\mathcal{A}}_{\scalebox{0.7}{D0-D4}}  &=&\hspace{-4mm} \sum_{p_i, s \in \{ 0,1,2\}}   \sum_{k,k^\prime\in\{0,1\}} \hspace{-2mm}   \frac{W_{2}^{[s]}}{\eta^4}  \bigl(\mathcal{V}  \xi^9_{2s  +6p_1} + \mathcal{H}  \xi^9_{3+2s  +6p_1} \bigr) \xi^9_{3k+2 s + 6 p_2} \xi^9_{3k^\prime+2 s + 6 p_3} \xi^3_{3k} \xi^3_{3k^\prime}  \; . 
%&=&\hspace{-1mm} \sum_{s \in \{ 0,1,2\}}   \sum_{k,k^\prime\in\{0,1\}} \hspace{-2mm}   \frac{W_{2}^{[s]}}{\eta^4}  \bigl(\mathcal{V}  \chi_{\frac{s}{3}} + \mathcal{H}  \chi_{\frac{1}{2} +\frac{s}{3} } \bigr) \chi_{\frac{k}{2}+\frac{ s}{3}  }  \chi_{\frac{k^\prime}{2}+\frac{ s}{3}  }   \xi^3_{3k} \xi^3_{3k^\prime} \; .  
\eea
The massless sector of the theory consists in a single $\mathcal{N}=4$ Maxwell multiplet, as expected for a fundamental brane. 
One obtains the tension square
  \be \widetilde{\mathcal{A}}^{{\rm NS}}_{\scalebox{0.7}{D0-D4}}|_{q=0} =    \mathcal{V}^{\rm NS}  \frac{P_{2}^{0\, {\rm mod}\, 3}}{T_2\eta^4} (\xi_0^9)^3 (\xi_0^3)^2 \Big|_{q=0}= \frac{4 }{T_2} \; , \ee
as for the $\mathds{Z}_2$ orbifold.

 \subsection{The non-geometric D2-D6 brane}

To obtain the transverse channel amplitude of a non-geometric D2-D6 brane, we need to find an isomorphism between the twisted sector and the untwisted sector that allows to define Fricke T-duality. This symmetry between the two sectors is not manifest in \eqref{TorusZ3}, but we are going to see that the isomorphism is an appropriate outer automorphism of $\mathcal{A}_9^3$. The 72 irreducible $ \mathcal{A}_9^3$ characters \eqref{A39Char} are not independent. There are related by the obvious automorphisms defined as permutations of the three $U(1)_1$ modules, and the automorphism $\varphi_i \rightarrow - \varphi_i$ that gives $(r,s)\rightarrow (-r,-s)$. 

Fricke T-duality involves yet another automorphism, that can be identified as  the field redefinition 
 \be \tilde{\varphi}_i = \frac{1}{3} \bigl( \varphi_i -2\varphi_{i+1} - 2 \varphi_{i+2}  \bigr) \; . \ee
 One checks that it preserves the algebra of the $U(1)_1$ currents because $(\frac{2}{3})^2 + (\frac{2}{3})^2+(\frac{1}{3})^2 = 1$ and  $\widetilde{\cal J}_i = i \partial \tilde{\varphi}_i$ are level 1 currents. It does not stabilise the subalgebra  $\mathcal{A}_9 \times \mathcal{A}_9 \times \mathcal{A}_9\subset \mathcal{A}_9^3 $ because 
 \be e^{\pm 6 i  \varphi_i} = e^{\mp 4 i ( \tilde{\varphi}_1 + \tilde{\varphi}_2+\tilde{\varphi}_3)} e^{\pm 6 i \tilde{\varphi}_i} \; , \ee
 but it does stabilise $ \mathcal{A}_9^3 $.  It acts on the irreducible  $ \mathcal{A}_9^3 $ modules of characters \eqref{A39Char} as $\xi_9^3[^{r,s}_{\; k_i}] \rightarrow \xi_9^3[^{s,r}_{\; k_i}]$. The character identity associated to this automorphism is given in \eqref{A39Identity}. A Fricke T-duality is defined to act on the momenta and winding as $m_2\rightarrow  3 n_2$ and $n_2\rightarrow  \frac{m_3}{3}$ \cite{ Bianchi:2022tbr}. One has therefore the symmetry of the partition function \eqref{TorusZ3} obtained by exchanging $(r,s)\rightarrow (s,r)$ on both the $T^2$ Narain partition function and the $\mathcal{A}_9^3 $ characters $\xi_9^3[^{r,s}_{\; k_i}] $. One can identify the  twisted sector $ \mathcal{A}_9^3 $ modules $\xi_9^3[^{0,s}_{\; k_i}] $ with the untwisted sector modules $\xi_9^3[^{s,0}_{\; k_i}] $. We show in Appendix \ref{SuperCurrent} that this isomorphism is consistent with worldsheet supersymmetry.

This identification allows to show explicitly that the trivial orbifold theory obtained by the asymmetric orbifold acting on $T^4$ without the circle translation is the maximally supersymmetric theory. In this case the component $\mathcal{V} \xi_9^3[^{\; 00}_{000}] (\xi_0^3)^2 / \eta^4$ (times the right-moving sector) includes the massless $\mathcal{N}=6$ gravity multiplet, while the two  $\mathcal{N}=6$ gravitini  multiplets come from the twisted sector  $\mathcal{H} \xi_9^3[^{\; 01}_{200}] (\xi_0^3)^2 / \eta^4$ and $\mathcal{H} \xi_9^3[^{\; 02}_{100}] (\xi_0^3)^2 / \eta^4$. The description of 1/2 BPS branes in this trivial asymmetric orbifold are `non-geometric' branes that couple to the twisted sector through the identification  associated to $\xi_9^3[^{0,s}_{\; k_i}] \cong \xi_9^3[^{s,0}_{\; k_i}]$ of the right-moving fermion field along $T^4$ with the product of the fermion and boson twisted fields.

 It is therefore meaningful to consider this possibility in the $\mathcal{N}=6$ orbifold theory. It follows indeed by Fricke T-duality that we can write the transverse channel amplitude for a non-geometric D2-D6 brane
\be  \widetilde{\mathcal{A}}_{\scalebox{0.7}{D2-D6}}  =\frac{1}{3}  \sum_{s=0}^2  \frac{T_2 W_{2}^{[s]}}{\eta^4}  \hspace{-4mm} \sum_{p,p^\prime\in \{ 0,1,2\}}  \hspace{-5mm} \sum_{\substack{ \ell_i \in \{ 0,1,2\} \\ \ell_1+\ell_2+\ell_3 = s \; {\rm mod}\;  3}}  \hspace{-3mm} \sum_{k,k^\prime\in\{0,1\}} \hspace{-2mm} ( \mathcal{V} \xi^9_{6\ell_1} + \mathcal{H} \xi^9_{6\ell_1+9})  \xi^9_{6\ell_2+9k}\xi^9_{6\ell_3+9k^\prime } \xi^3_{2p+3k}\xi^3_{2p^\prime+3k^\prime} \; .  \ee
The direct channel amplitude then reads 
\be  {\mathcal{A}}_{\scalebox{0.7}{D2-D6}}  = \sum_{p_i, \ell \in \{ 0,1,2\}}   \sum_{k,k^\prime\in\{0,1\}} \hspace{-2mm}   \frac{P_{2}^{ \ell \; {\rm mod}\; 3}}{\eta^4}  \bigl(\mathcal{V}  \xi^9_{2\ell +6p_1} + \mathcal{H}  \xi^9_{3+2\ell +6p_1} \bigr) \xi^9_{3k+2 \ell + 6 p_2} \xi^9_{3k^\prime+2 \ell + 6 p_3} \xi^3_{3k} \xi^3_{3k^\prime} \; . \ee
%\bea  {\mathcal{A}}_{\scalebox{0.7}{D2-D6}}  &=&\hspace{-4mm} \sum_{p_i, \ell \in \{ 0,1,2\}}   \sum_{k,k^\prime\in\{0,1\}} \hspace{-2mm}   \frac{P_{2}^{ \ell \; {\rm mod}\; 3}}{\eta^4}  \bigl(\mathcal{V}  \xi^9_{2\ell +6p_1} + \mathcal{H}  \xi^9_{3+2\ell +6p_1} \bigr) \xi^9_{3k+2 \ell + 6 p_2} \xi^9_{3k^\prime+2 \ell + 6 p_3} \xi^3_{3k} \xi^3_{3k^\prime}   %\CR
%%&=&\hspace{-1mm} \sum_{\ell \in \{ 0,1,2\}}   \sum_{k,k^\prime\in\{0,1\}} \hspace{-2mm}   \frac{P_{2,2}^{\ell \; {\rm mod}\; 3}}{\eta^4}  \bigl(\mathcal{V}  \chi_{\frac{\ell}{3}} + \mathcal{H}  \chi_{\frac{1}{2} +\frac{\ell}{3} } \bigr) \chi_{\frac{k}{2}+\frac{ \ell}{3}  }  \chi_{\frac{k^\prime}{2}+\frac{ \ell}{3}  }   \xi^3_{3k} \xi^3_{3k^\prime}  
%\eea
The massless content of the theory is just one $\mathcal{N}=4$ Maxwell multiplet, as for the $\mathds{Z}_2$ orbifold. The tension square of the non-geometric  brane is 
  \be \widetilde{\mathcal{A}}^{{\rm NS}}_{\scalebox{0.7}{D2-D6}}|_{q=0} =   \frac13 \mathcal{V}^{\rm NS}  \frac{T_2 W_{2}^{[0]}}{\eta^4} (\xi_0^9)^3 (\xi_0^3)^2 \Big|_{q=0}= \frac{4 T_2 }{3} \; , \ee
so that the tension has the expected additional factor of  $\frac{1}{\sqrt{3}} $ predicted by Fricke T-duality \cite{ Bianchi:2022tbr}. 

As for the $\mathds{Z}_2$ orbifold, one may argue that a geometric brane can be obtained as a D2-D4-D6 bound state of a D2-D6 brane and two D4-D4 branes. Here we do not discuss in details the relation between the RR charges and the geometric configuration of branes, but we want to stress that the geometric brane and the non-geometric brane cannot have the same RR charge up to a factor of $\sqrt{3}$, because this is inconsistent with the lattice of charges $\mathds{H}(3)= A_2 \oplus A_2 $ \cite{Bianchi:2022tbr}.\footnote{An example is to take the non-geometric charges $\frac{e_3}{\sqrt{3}}$ and $\frac{1}{2} - \frac{e_3}{2\sqrt{3}}$ and the geometric charge $1 = \frac{e_3}{\sqrt{3}}+2\times (\frac{1}{2} - \frac{e_3}{2\sqrt{3}})$ with $e_3$ a unit imaginary quaternion.}

 The transverse channel amplitude is simply obtained by assuming that it does not couple to the twisted sector and one gets 
\be  \widetilde{\mathcal{A}}_{\scalebox{0.7}{D2-D4-D6}}  =  \frac{T_2 W_{2}^{[0]}}{\eta^4}  \hspace{-4mm} \sum_{p,p^\prime\in \{ 0,1,2\}}  \hspace{-5mm} \sum_{\substack{ \ell_i \in \{ 0,1,2\} \\ \ell_1+\ell_2+\ell_3 = 0 \; {\rm mod}\;  3}}  \hspace{-3mm} \sum_{k,k^\prime\in\{0,1\}} \hspace{-2mm} ( \mathcal{V} \xi^9_{6\ell_1} + \mathcal{H} \xi^9_{6\ell_1+9})  \xi^9_{6\ell_2+9k}\xi^9_{6\ell_3+9k^\prime } \xi^3_{2p+3k}\xi^3_{2p^\prime+3k^\prime} \; ,  \ee
and the direct channel amplitude gives 
\be  {\mathcal{A}}_{\scalebox{0.7}{D2-D4-D6}}  = \sum_{p_i, \ell \in \{ 0,1,2\}}   \sum_{k,k^\prime\in\{0,1\}}   \frac{P_{2}}{\eta^4}  \bigl(\mathcal{V}  \xi^9_{2\ell +6p_1} + \mathcal{H}  \xi^9_{3+2\ell +6p_1} \bigr) \xi^9_{3k+2 \ell + 6 p_2} \xi^9_{3k^\prime+2 \ell + 6 p_3} \xi^3_{3k} \xi^3_{3k^\prime} \; . \ee
In this case the massless sector includes one $\mathcal{N}=4$ vector multiplet that comes from $( \mathcal{V}  \xi^9_{0} + \mathcal{H}  \xi^9_{3}+ \mathcal{H}  \xi^9_{-3} \bigr) (\xi^9_{0})^2 ( \xi^3_{0})^2$ and one $\mathcal{N}=2$ hyper-multiplet from $ \mathcal{H}  \xi^9_{-1} (\xi^9_{2})^2 ( \xi^3_{0})^2+ \mathcal{H}  \xi^9_{1} (\xi^9_{-2})^2 ( \xi^3_{0})^2$.  The tension square of the geometric brane is 
  \be \widetilde{\mathcal{A}}^{{\rm NS}}_{\scalebox{0.7}{D2-D4-D6}}|_{q=0} =   \mathcal{V}^{\rm NS}  \frac{T_2 W_{2}^{[0]}}{\eta^4} (\xi_0^9)^3 (\xi_0^3)^2 \Big|_{q=0}=4 T_2 \; , \ee
so one finds a factor of $\sqrt{3}$ between the masses of the geometric branes and the non-geometric brane, as expected from Fricke T-duality  \cite{Bianchi:2022tbr}.

\section{Conclusions and outlook}

We have given a detailed (microscopic) description of the geometric and non-geometric fundamental branes and their bound states in two classes of T-fold backgrounds with ${\cal N}=6$ supersymmetry. As shown in \cite{Bianchi:2022tbr} these correspond to $\mathds{Z}_2$ and $\mathds{Z}_3$ asymmetric orbifolds that combine a non-geometric `T-duality' action on $T^4$ with a shift along $T^2$. 

Focussing on the Type IIA framework we have first analyzed the $\mathds{Z}_2$ case and identified both geometric and non-geometric fundamental branes. After discussing D0-D4 whose tension is the one expected for a geometric brane, we passed to consider a D0-D2-D4 boundary state, whose  direct construction leads to a supersymmery breaking brane with a scalar tachyon. We argue by Fricke duality that tachyon condensation  leads to a supersymmetric bound state with tension  equal to $\sqrt{2}$ times the one of the geometric  D0-D4 brane. This bound state is not geometric in the sense that the Ishibashi state combine world-sheet fermions and bosons $SU(2)_1$ modules. The same construction holds for the non-geometric D2-D6 brane, which has tension equal to $1/\sqrt{2}$ the one of the geometric D2-D4-D6 brane.

In the $\mathds{Z}_3$ case we found similar results for geometric D0-D4 brane and for  non-geometric D2-D6 brane.  The non-geometric brane involves again a mixing between world-sheet fermions and bosons. One finds a factor of $\sqrt{3}$ between the tension of geometric and non-geometric branes, as expected from Fricke T-duality. 

It would interesting to extend our analysis to other T-fold backgrounds with lower or no supersymmetry \cite{Anastasopoulos:2009kj, Bianchi:2012xz} or to S-folds \cite{Garcia-Etxebarria:2015wns, Heckman:2020svr, Giacomelli:2023qyc} or U-folds \cite{Braun:2013yla, Candelas:2014jma, Candelas:2014kma}. The price one has to pay is the lack of a `perturbative' world-sheet description that in the present case proved crucial in checking consistency between open- and closed-string descriptions. Knowledge of the relevant U-duality may be sufficient  to determine the lattice of charges as in the ${\cal N}=6$ case \cite{Bianchi:2022tbr}

Constructions of more general bound states using magnetised D0-D4 branes  or T-branes \cite{Collinucci:2014qfa, Bena:2019rth}, that could lead to supersymmetry enhancement, may well provide new useful insights in this endeavour.

%%%%%%%%%%%%%%%%%%%%%%%%
\subsection*{Acknowledgements}

We would like to thank Carlo Angelantonj, Costas Bachas, Emilian Dudas, Matthias Gaberdiel and Gianfranco Pradisi for useful discussions. MB thanks the MIUR PRIN contract 2020KR4KN2 ``String Theory as a bridge between Gauge Theories and Quantum Gravity'' and the INFN project ST$\&$FI ``String Theory and Fundamental Interactions'' for partial support.  MB gratefully acknowledges the hospitality of CPHT and CNRS while this project was started.

\appendix

\section*{Appendix}

 \section{R-R charges and rank reduction}
\label{RRcharges} 
We use the convention that the left and right projections of the string zero modes of momentum $\vec{m}$ and winding number $\vec{n}$ along $T^4$ read 
\bea p_L(\vec{m},\vec{n})^2 &=& \frac{1}{2} G^{-1} ( \vec{m} + (G + B) \vec{n} , \vec{m} + (G + B) \vec{n}) \; , \CR
 p_R(\vec{m},\vec{n})^2 &=& \frac{1}{2} G^{-1} ( \vec{m} + (-G + B) \vec{n} , \vec{m} + (-G + B) \vec{n}) \; . \label{MomentaLeftRight} \eea
Recall that we define the left and right momenta without dimensions, such that they appear as $e^{\pi i \tau p_L^2 - \pi i \bar \tau p_R^2}$ in the Narain partition function. A $D_4$ symmetric point is obtained at 
\be G = \frac{1}{2} A = \left( \begin{array}{cccc} 1 \; & - \frac12 \; &\; 0\; &\; 0\\ - \frac12 \; & \; 1\; &-\frac12\; & - \frac12 \\
0\; &- \frac12 & \; 1\; &\; 0\\ 0\; &- \frac12 & \; 0 \; &\; 1\end{array}\right)\; , \qquad B = \left( \begin{array}{cccc} 0\; & \;  \frac12 \; &\; 0\; &\; 0\\ - \frac12 \; & \; 0\; &\; \frac12\; & \;  \frac12 \\
0\; &- \frac12 & \; 0\; &\; 0\\ 0\; &- \frac12 & \; 0 \; &\; 0\end{array}\right)  \; , \label{GBZ2}\ee
where $A$ is the $D_4$ Cartan matrix and $G + B \in SL(4,\mathds{Z})$. 

The identification with the lattice can be made explicit with 
\be p_L = \frac1{2} G^{-1} ( \vec{m} + (G + B) \vec{n} ) \; , \qquad p_R = \frac1{2} G^{-1} ( \vec{m} + (-G + B) \vec{n} ) \ee
such that 
\be p_L^2  = A( p_L,p_L) \; , \qquad p_R^2 = A( p_R,p_R)\; , \ee
and $(p_L,p_R) \in (D_4 + \mu)\otimes (D_4 + \mu)$ for $\mu \in D_4^*/ D_4$. 

One easily checks that $(G-B)^{-1}  (G+B) $ preserves the weight 
\be (G-B)^{-1}  (G+B)  (D_4+\mu) = D_4+\mu \; , \ee
therefore for any Weyl reflection $w\in W(D_4)$, the structure is preserved and one can define the boundary condition  \footnote{$w=1$ is Neumann, $w=-1$ Dirichlet, and other Weyl reflections should correspond to mixed boundary conditions.}
\be  p_L = \pm w  (G-B)^{-1}  (G+B) p_R  \in D_4+\mu \; , \ee  
for each right-moving state  $p_R \in D_4+\mu$, with the sign depending of the eigen value of the state under $\mathds{Z}_2$.

The D-brane charges along $T^4$ are the D0 charge $q\in \mathds{Z}$, the D2 charge  $Q^{ab} \in \wedge^2 \mathds{Z}^4$, and the D4 brane charge $p\in \mathds{Z}$, with the `left' and `right' projections 
\bea  p_L^S(q,Q,p)^2 &=& \frac{1}{2 |G|^\frac12} \bigl( q- \tfrac12 \tr[ B Q]  - \tfrac14 \tr[ B^\star B] p  - |G|^\frac12 p\bigr)^2\CR
&& \qquad - \frac14 \tr\bigl[ ( Q + B^\star p) \cdot \bigl(  |G|^{-\frac12} G (Q+B^\star p ) G +( Q+B^\star p)^\star \bigr)\bigr] \CR
p_R^S(q,Q,p)^2 &=& \frac{1}{2 |G|^\frac12} \bigl( q- \tfrac12 \tr[ B Q]  - \tfrac14 \tr[ B^\star B] p  + |G|^\frac12 p\bigr)^2\CR
&& \qquad - \frac14 \tr\bigl[ ( Q + B^\star p) \cdot \bigl(  |G|^{-\frac12} G (Q+B^\star p ) G -( Q+B^\star p)^\star \bigr)\bigr] \; , \label{PSLR} \eea 
where $|G| = \det G$ and $B^\star_{ab} = \frac{1}2 \varepsilon_{abcd} B^{cd}$. The $\mathds{Z}_2$ orbifold condition $p_L^S(q,Q,p)=0$ can be solved for integral coefficients at the $D_4$ symmetric point and for the appropriate parametrisation 
\be q = q_1 + q_2 + q_4\, , \quad Q =  \left( \begin{array}{cccc} 0\; &\;   - q_2 -\sum_{i=1}^4 q_i\; &\; -q_1-q_2\; &\; -\sum_{i=1}^4 q_i \\  q_2 +\sum_{i=1}^4 q_i \; & \; 0\; &\; q_3\; & -q_1 \\
q_1+q_2 \; &- q_3 & \; 0\; &\; q_2\\ \sum_{i=1}^4 q_i \; &q_1 & -q_2   \; &\; 0\end{array}\right)  \; ,  \quad p = q_4\; , \label{DbraneChargesT4} \ee
one obtains
\be p_R^S(q)^2 = 2 A^{-1}(\vec{q},\vec{q})\; . \ee
This shows that the branes localised in the twisted circle, combining a brane and its image under the $\mathds{Z}_2$ action, have R-R charges in $ \mathds{H}$. 

Because of the Chern-Simons coupling $\int_{D4} e^{B} C$ on the D-brane world volume, the R-R charge induced by $N_4$ D4-branes and $N_0$ D0-branes in the presence of a constant $B$ field is  
\be q =N_0-\tfrac14 \tr[ B^\star B]  N_4 \; , \qquad   Q = - B^\star N_4 \; , \qquad p = N_4\; . \ee 
In order to get a R-R charge in $\mathds{H}$ and invariant under $\mathds{Z}_2$, one must take $N_4=2 N_0$ even, with $q_2 = -N_0$ and $q_4=2N_0$. One then gets the RR charge $q=  N_0\in \mathds{H}$.  Since one can define the boundary condition for any Weyl group element, one find identically that $q = N_0 u_I \in \mathds{H}$ for any unit $u_I$.  

Because $B$ is rank one and half-integral one expects the effective theory on the D4 brane to be non-commutative, with in the appropriate basis for the coordinates 
\be \{ X^4 , X^i \} = 0 \; , \qquad [ X^i , X^j ] = 0 \; , \ee
with $i=1,3$. This has the solution 
\be X^4 =x^4  \sigma_3 \; , \qquad X^i = x^i  \sigma_1 \ee
where $x^0$ and $x^i$ are real valued coordinates on the torus.  The effective rank of the gauge theory is therefore reduced by half. This seems consistent with the property that the $\mathds{Z}_2$ action exchanges $N_0$ and $\frac{N_4}{2}$ when acting on the R-R charges.

\vskip 4mm

One can do a similar computation for the $\mathds{Z}_3$ orbifold at the $A_2\oplus A_2$ symmetric point 
\be G = \left( \begin{array}{cccc} 1 \; & - \frac12 \; &\; 0\; &\; 0\\ - \frac12 \; & \; 1\; &\; 0\; &\; 0  \\
0\; &\; 0  & \; 1\; &- \frac12 \\ 0\; &\; 0  & - \frac12  \; &\; 1\end{array}\right)\; , \qquad B = \left( \begin{array}{cccc} 0\; & \;  \frac12 \; &\; 0\; &\; 0\\ - \frac12 \; & \; 0\; &\; 0\; & \;  0 \\
0\; &\; 0  & \; 0\; &\;  \frac12 \\ 0\; &\; 0  & - \frac12 \; &\; 0\end{array}\right)  \; , \label{A2A2} \ee
where again
\be p_L = \frac1{2} G^{-1} ( \vec{m} + (G + B) \vec{n} ) \; , \qquad p_R = \frac1{2} G^{-1} ( \vec{m} + (-G + B) \vec{n} ) \ee
such that 
\be p_L^2  = A( p_L,p_L) \; , \qquad p_R^2 = A( p_R,p_R)\; , \ee
and $(p_L,p_R) \in (A_2\oplus A_2 + \mu)\otimes (A_2\oplus A_2 + \mu)$ for $\mu \in (A_2^*/ A_2)^2$. In this case one checks that $(G-B)^{-1}  (G+B) $ preserves the weight 
\be (G-B)^{-1}  (G+B)  (A_2\oplus A_2+\mu) = A_2\oplus A_2-\mu \; , \ee
therefore for any Weyl reflection $w\in W(A_2\oplus A_2)$, the structure is preserved and one can define the boundary condition
\be  p_L = \omega^n w  (G-B)^{-1}  (G+B) p_R  \in A_2\oplus A_2-  \omega^n w  \mu \; , \ee  
for each right-moving state  $p_R \in A_2\oplus A_2+\mu$, with $\omega$ the Weyl group element identified with the action of $\mathds{Z}_3$ and $n=0,1,2$ depending of the eigen value of the state under $\mathds{Z}_3$.

One concludes as for $D_4= \mathds{H}(2)$  that the  D-brane charges are valued in $ A_2 \oplus A_2=\mathds{H}(3) $. Then a combinations of $N_4$ D4 branes and $N_0$ D0 branes gives an integral R-R charge in $ \mathds{H}(3)$ if $N_4= 0$ mod $4$, and $N_0 = \frac{3}{4} N_4$. One then gets $q=  N_4/4\in  \mathds{H}(3)$.  

Having a multiple of four can be justified since the $B$-field is now rank $4$ and one gets the non-commutative coordinates 
\be \{ X^\alpha , X^{\beta} \} = 0 \; , \qquad [X^\alpha , X^{\hat{\beta}} ] = 0 \; , \qquad  \{ X^{\hat{\alpha}} , X^{\hat{\beta}} \} = 0 \; ,  \ee
which requires to reduce the rank by four with
\be X^\alpha = x^\alpha \sigma_\alpha \otimes \mathds{1}  \; , \quad X^{\hat{\alpha}} =  x^{\hat{\alpha}}  \mathds{1}\otimes  \sigma_{\hat{\alpha}}  \; . \ee
The action of  $\mathds{Z}_3$ on the charges is more subtle in this case, since one needs to include D2 branes to get a triplet. 

\section{Affine characters}
In this appendix we give our notations for affine level one characters. We use the $SO(2n)_1$ characters 
\be 
O_{2n} = \frac{\vartheta_3^n + \vartheta_4^n }{2 \eta^n} \, , \quad 
V_{2n} = \frac{\vartheta_3^n - \vartheta_4^n }{ 2 \eta^n} \, ,\quad
S_{2n} = \frac{\vartheta_2^n + i^n \vartheta_1^n }{ 2 \eta^n} \, ,\quad
C_{2n} = \frac{\vartheta_2^n - i^n\vartheta_1^n }{ 2 \eta^n} \; . \label{SO2nChar} 
\ee
where $\vartheta_\alpha$ are Jacobi (elliptic) theta functions. We also use the $SU(2)_1$ characters 
 \be \chi_j(z,\tau)  = \frac{1}{\eta(\tau)} \sum_{n\in \mathds{Z}} e^{2\pi i z (n+j)} q^{(n+j)^2}   \; , \label{SU2Char} \ee
with $j=0$ or $\frac{1}{2}$. In particular we have 
\be O_4 = \chi_0 \chi_0\; , \qquad V_4 = \chi_\frac12 \chi_\frac12 \; , \qquad S_4 = \chi_\frac12 \chi_0\; , \qquad C_4 =   \chi_0 \chi_\frac12  \; . \label{OVCS4}  \ee
We write the $U(1)_1$ characters 
\be \xi^N_p(y) = {\rm Tr}_{\mathcal{H}_p} \,  q^{L_0- \frac{1}{24}} y^{J_0} = \frac{\sum_{n} q^{N(n+ \frac{p}{2N})^2}y^{N n +\frac{p}{2}}}{\eta}  \; , \label{U1Char}  \ee
that satisfy 
\bea \xi^N_p(- \tfrac1 \tau)  &= & \frac{1}{\sqrt{2N}} \sum_{q=0}^{2N-1}  e^{- \pi i \frac{ pq}{N}} \xi^N_q(\tau) \; , \CR
\xi^N_p(\tau+1)  &=& e^{\pi i ( \frac{p^2}{2N}- \frac{1}{12})} \xi^N_p(\tau) \; . 
\eea
One can decompose the $SU(2)_1$ characters into $U(1)_1$ characters 
\be \chi_j(z) = \sum_{\ell=0}^{N-1} \xi_{2N(\ell+j)}^{N^2}( y = e^{\frac{2\pi i z}{N}}) \; , \ee
for any integers $N$, which gives in particular for any integer $p$
\be \chi_j(\tfrac{p}{N}) = \sum_{\ell=0}^{N-1} e^{\frac{2\pi i p }{N}(\ell+j)} \xi_{2N(\ell+j)}^{N^2}\; . \ee
By inversion one gets 
\be \chi_{\frac{p}{N}} = \sum_{\ell=0}^{N-1} \xi_{2(N\ell+p)}^{N^2}\; , \label{SU2U1}  \ee
which is not an $SU(2)_1$ character but the sum of $N$ $U(1)_1$ characters. 

The $T^4$ mode number partition function with the insertion of the $\mathds{Z}_3$ operator can be written in terms of $U(1)_1$ characters as follows 
 \be \frac{\eta}{ i q^\frac{1}{18} \vartheta_1(\frac{\tau}{3})} =  ( \xi_{4s}^9 + \xi_{4s+6}^9 +  \xi_{4s+12}^9  ) \xi_{2p}^3 + (  \xi_{4s+9}^9 +  \xi_{4s-3}^9 + \xi_{4s+3}^9 )   \xi_{2p+3}^3 \label{thetaxixi} \ee
 for $s=1,2$ and all $p=0,1,2$, so in particular  one has 
 \be \xi^9_{4s+6\ell } \xi^3_0 +  \xi^9_{4s+6\ell +9} \xi^3_3 = \xi^9_{4s + 6(\ell-s) } \xi^3_{2}  +  \xi^9_{4s + 6(\ell-s) +9} \xi^3_{5} = \xi^9_{4s + 6(\ell-s) } \xi^3_{4}  +  \xi^9_{4s + 6(\ell-s) +9} \xi^3_{1} \; . \ee

The  $\mathcal{A}_9^3$ characters \eqref{A39Char} transform under modular inversion as
\be \xi_9^3[^{r,s}_{\; k_i}](- \tfrac1{\tau}) = \frac{1}{6\sqrt{2}} \sum_{r^\prime,s^\prime \in \{ 0,1,2\} } \sum_{k^\prime_i \in \{ 0,1\}} \omega^{r s^\prime + s r^\prime + s s^\prime} (-1)^{\sum_i k_i k_i^\prime} \xi_9^3[^{r^\prime ,s^\prime }_{\; k^\prime_i}](\tau)\; , \ee
and they satisfy $ \xi_9^3[^{r,s}_{\; k_i}] =  \xi_9^3[^{s,r}_{\; k_i}]$ as can be checked from the identity 
\bea && \sum_{\ell_1+\ell_2+\ell_3 +r+s= 0 \; {\rm mod}\;  3} \xi^9_{4s+6\ell_1+9k_1 }(y_1)  \xi^9_{4s+6\ell_2+9k_2 }(y_2) \xi^9_{4s+6\ell_3+9k_3 } (y_3) \label{A39Identity} \\
 &=&  \sum_{\ell_1+\ell_2+\ell_3 +r+s = 0 \; {\rm mod}\;  3} \xi^9_{4r+6\ell_1+9k_1 } \Bigl( \bigl(\tfrac{y_1}{y_2^2 y_3^2}\bigr)^{\frac13}  \Bigr)  \xi^9_{4r+6\ell_2+9k_2 }\Bigl( \bigl(\tfrac{y_2}{y_3^2 y_1^2}\bigr)^{\frac13}  \Bigr) \xi^9_{4r+6\ell_3+9k_3 } \Bigl( \bigl(\tfrac{y_3}{y_1^2 y_2^2}\bigr)^{\frac13}  \Bigr)\; . \nonumber \eea
 for any integers $k_i$ and $r,s$. 
 
\section{Supersymmetry current} 
\label{SuperCurrent}
In this appendix we show that the supersymmetry current is invariant under Fricke duality.\footnote{We would like to thank C.~Bachas for suggesting us to clarify the consistency of the isomorphism exploited in the construction of non-geometric branes with the $\mathcal{N}=1$ worldsheet supersymmetry current.} For this we use a representation of all fields associated to $T^4$ in terms of free bosons \cite{Lerche:1988np}, such that the $T^4$ supersymmetry current is  
\be G(z) = \frac{1}{2\sin(\frac{\pi}{K})} \sum_{\lambda \in \Delta} c_\lambda  \; e^{i (\lambda , \phi(z))}   \; , \ee
where $\phi(z)$ is a vector of six free bosons and $\Delta$ is a set of $16 \sin(\frac{\pi}{K})^2$ vectors of norm square $(\lambda,\lambda)=6$, that is invariant under the $-\mathds{1}$ reflection  and a Fricke reflection. This gives  $16$ vectors for $\mathds{Z}_2$ and  $12$ for $\mathds{Z}_3$.  The cocycle $c_\lambda$ is determined such that $G(z)$ is a fermion field and its operator product expansion gives the singular terms  
\be G(z) G(0) \sim \frac{4}{z^3} - \frac{2}{z} \partial \phi \cdot \partial \phi  \; . \label{OPEGGT} \ee
This requires in particular the conditions 
\be \sum_{\lambda \in \Delta} \lambda = 0 \; , \qquad  \frac{1}{16 \sin(\frac{\pi}{K})^2} \sum_{\lambda \in \Delta} \lambda \otimes \lambda =  \mathds{1}\; . \ee
To compute the cocycle it is convenient to introduce the $T^4$ bosons and fermions  
\be  G(z) = \psi_{\alpha\dot{\beta}}(z) \partial X^{\alpha\dot{\beta}}(z) \; . \ee
We write $\sigma$ for the R-symmetry $SU(2)$ and $\varphi_\psi$ for the broken $SU(2)$ such that the complex fermions take the form 
\be \psi_{\alpha\dot{\beta}} = c_{\alpha,\dot{\beta}} \, e^{i \alpha \sigma + i \dot{\beta} \varphi_\psi} \; , \ee
for $\alpha,\dot{\beta} = \pm 1$. We can choose the cocycle $c_{\alpha,\dot{\beta}}$ such that $c_{++}=c_{--}=1$ and $c_{+-} = -c_{-+}$ satisfies 
\be c_{+-} e^{i \alpha \sigma } = (-1)^{\alpha}  e^{i \alpha \sigma } c_{+-} \; . \label{cocycleFermions} \ee
\subsection{$SU(2)^6$}
For the $\mathds{Z}_2$ orbifold one introduces a free boson for each $SU(2)$ affine algebra. We define
\be \partial X^{+\pm } =  \frac{1}{2} \Bigl( e^{i (\pm  \varphi_o + \varphi_v+\varphi_s + \varphi_c)}  \pm e^{i (\pm \varphi_o + \varphi_v-\varphi_s - \varphi_c)}   + c_v e^{i (\mp \varphi_o - \varphi_v-\varphi_s +\varphi_c)}\pm  c_v e^{i (\mp \varphi_o - \varphi_v+\varphi_s -\varphi_c)} \Bigr) \ee
with the cocycle $c_v$ satisfying 
\be c_{v} e^{i \alpha \varphi_v } = (-1)^{\alpha}  e^{i \alpha \varphi_v } c_{v} \; . \label{cocycleV} \ee
The other bosons are obtained by conjugation as
\be \partial X^{-\pm } =  \frac{1}{2} \Bigl( e^{i (\mp   \varphi_o - \varphi_v-\varphi_s - \varphi_c)}  \pm e^{i (\mp \varphi_o - \varphi_v+\varphi_s + \varphi_c)}   - c_v e^{i (\pm  \varphi_o +\varphi_v+\varphi_s -\varphi_c)}\mp   c_v e^{i (\pm \varphi_o +\varphi_v-\varphi_s +\varphi_c)} \Bigr) \; . \ee

With the convention that $\phi$'s components are   
\be \phi = ( \sigma , \varphi_\psi, \varphi_o , \varphi_v,\varphi_s , \varphi_c) \; , \ee
we write the vectors $\lambda$ as
\be \lambda = ( \beta \gamma \delta , \alpha , \alpha \gamma \delta , \beta, \gamma , \delta ) \; , \ee
for $\alpha,\beta,\gamma,\delta = \pm 1$. One checks that this set is invariant under reflection by changing the sign of all $\alpha,\beta,\gamma,\delta$. It is invariant under the exchange of the second  and the third component, i.e. Fricke reflection, through the redefinition $\alpha \rightarrow \alpha \gamma \delta$. The cocycle $c_\lambda$ is not invariant under the Fricke reflection, but the new cocycle is by construction consistent with the operator product expansion \eqref{OPEGGT} and their difference can be absorbed in trivial phases redefinitions of the momentum eigenspaces for the six bosons.

\subsection{$SU(2) \times U(1)^5$}
For the $\mathds{Z}_3$ orbifold we use the convention
\be \phi = ( \sigma , \varphi_1, \varphi_2 , \sigma_2 ,\varphi_3 , \sigma_3) \; , \ee
where we introduced $\varphi_1=\varphi_\psi$ for convenience and the twelve vectors $\lambda$ are the rows of the matrix
\be \lambda = \left(\begin{array}{cccccc} 1&  1& -1 & \sqrt{3} &0&  0 \\ 
-1 &  1 &  0&  0& -1& \sqrt{3} \\
 1&  1& -1& -\sqrt{3}& 0& 0 \\
  -1& 1& 0& 0& -1& -\sqrt{3} \\
  -1& -1& 1& \sqrt{3}& 0& 0 \\
  1& -1& 0& 0& 1& \sqrt{3} \\
  -1& -1& 1& -\sqrt{3}& 0&  0 \\
  1& -1& 0& 0& 1& -\sqrt{3} \\ 1& 1& 2& 0& 0& 0 \\
  -1& 1& 0& 0& 2& 0 \\ -1& -1& -2& 0& 0& 0 \\ 1& -1& 0& 0& -2& 0 \end{array}\right) \; . 
\ee
This is consistent with the $\mathds{Z}_3$ orbifold action because $\lambda_2+\lambda_3 + \lambda_5 = 0 $ mod $3$ for all vectors. 
One defines accordingly the complex boson 
\be \partial X^{++}= \frac{1}{\sqrt{3}} \Bigl( e^{- i \varphi_2 + i \sqrt{3} \sigma_2 } + c_2 e^{2 i \varphi_2  } + c_2 e^{- i \varphi_2 - i \sqrt{3} \sigma_2 } \Bigr) \ee
with $c_2$ defined such that 
\be c_2 e^{\alpha i \sqrt{3} \sigma_2} = (-1)^\alpha  e^{\alpha i \sqrt{3} \sigma_2}c_2\; , \ee
and idem for $\partial X^{+-}$ with $2$ replaced by $3$.

One checks that this set of vectors $\lambda$ is invariant under the $- \mathds{1}$ reflection and under the Fricke reflection 
\be \varphi_i \rightarrow  \frac{1}{3} \bigl( \varphi_i -2 \varphi_{i+1}- 2\varphi_{i+2} \bigr) \; . \ee
The cocycle is not invariant, but this can be compensated by an appropriate set of phases.


\begin{thebibliography}{99}

\bibitem{Hull:2004in}
C.~M.~Hull,
``A Geometry for non-geometric string backgrounds,''
\doi{JHEP \textbf{10} (2005), 065}{doi:10.1088/1126-6708/2005/10/065}
\eprint{hep-th/0406102}.
%524 citations counted in INSPIRE as of 24 Mar 2022


\bibitem{Hull:1994ys}
C.~M. Hull and P.~K. Townsend, ``Unity of superstring dualities,'' {\em Nucl.
  Phys.} {\bf B438} (1995)  109--137,
\eprint{hep-th/9410167}.
%%CITATION = HEP-TH 9410167;%%.

%\cite{Narain:1986qm}
\bibitem{Narain:1986qm}
K.~S.~Narain, M.~H.~Sarmadi and C.~Vafa,
``Asymmetric Orbifolds,''
\doi{Nucl. Phys. B \textbf{288} (1987), 551.}{doi:10.1016/0550-3213(87)90228-8}
%489 citations counted in INSPIRE as of 24 Mar 2022

  %\cite{Ferrara:1989nm}
\bibitem{Ferrara:1989nm}
S.~Ferrara and C.~Kounnas,
``Extended supersymmetry in four-dimensional Type {II} Strings,''
\doi{Nucl. Phys. B \textbf{328} (1989), 406-438.}{doi:10.1016/0550-3213(89)90335-0}
%67 citations counted in INSPIRE as of 17 Jan 2022


%\cite{Narain:1990mw}
\bibitem{Narain:1990mw}
K.~S.~Narain, M.~H.~Sarmadi and C.~Vafa,
``Asymmetric orbifolds: Path integral and operator formulations,''
\doi{Nucl. Phys. B \textbf{356} (1991), 163-207.}{doi:10.1016/0550-3213(91)90145-N}
%88 citations counted in INSPIRE as of 24 Mar 2022


%\cite{Anastasopoulos:2009kj}
\bibitem{Anastasopoulos:2009kj}
P.~Anastasopoulos, M.~Bianchi, J.~F.~Morales and G.~Pradisi,
``(Unoriented) T-folds with few T's,''
\doi{JHEP \textbf{06} (2009), 032}{doi:10.1088/1126-6708/2009/06/032}
\eprintN{0901.0113}.
%10 citations counted in INSPIRE as of 26 Mar 2022

%\cite{Bianchi:2012xz}
\bibitem{Bianchi:2012xz}
M.~Bianchi, G.~Pradisi, C.~Timirgaziu and L.~Tripodi,
``Heterotic T-folds with a small number of neutral moduli,''
\doi{JHEP \textbf{10} (2012), 089}{doi:10.1007/JHEP10(2012)089}
\eprintN{1207.2665}.
%3 citations counted in INSPIRE as of 26 Mar 2022


%\cite{Braun:2013yla}
\bibitem{Braun:2013yla}
A.~P.~Braun, F.~Fucito and J.~F.~Morales,
``U-folds as K3 fibrations,''
\doi{JHEP \textbf{10} (2013), 154}{doi:10.1007/JHEP10(2013)154}
\eprintN{1308.0553}.
%22 citations counted in INSPIRE as of 01 Oct 2023

%\cite{Candelas:2014jma}
\bibitem{Candelas:2014jma}
P.~Candelas, A.~Constantin, C.~Damian, M.~Larfors and J.~F.~Morales,
``Type IIB flux vacua from G-theory I,''
\doi{JHEP \textbf{02} (2015), 187}{doi:10.1007/JHEP02(2015)187}
\eprintN{1411.4785}.
%23 citations counted in INSPIRE as of 01 Oct 2023

%\cite{Candelas:2014kma}
\bibitem{Candelas:2014kma}
P.~Candelas, A.~Constantin, C.~Damian, M.~Larfors and J.~F.~Morales,
``Type IIB flux vacua from G-theory II,''
\doi{JHEP \textbf{02} (2015), 188}{doi:10.1007/JHEP02(2015)188}
\eprintN{1411.4786}.
%22 citations counted in INSPIRE as of 01 Oct 2023

%\cite{Garcia-Etxebarria:2015wns}
\bibitem{Garcia-Etxebarria:2015wns}
I.~Garc\'\i{}a-Etxebarria and D.~Regalado,
``$ \mathcal{N}=3 $ four dimensional field theories,''
\doi{JHEP \textbf{03} (2016), 083}{doi:10.1007/JHEP03(2016)083}
\eprintN{1512.06434}.
%148 citations counted in INSPIRE as of 01 Oct 2023

%\cite{Heckman:2020svr}
\bibitem{Heckman:2020svr}
J.~J.~Heckman, C.~Lawrie, T.~B.~Rochais, H.~Y.~Zhang and G.~Zoccarato,
``$S$-folds, string junctions, and $\mathcal{N} = 2$ SCFTs,''
\doi{Phys. Rev. D \textbf{103} (2021) no.8, 086013}{doi:10.1103/PhysRevD.103.086013}
\eprintN{2009.10090}.
%25 citations counted in INSPIRE as of 01 Oct 2023

%\cite{Giacomelli:2023qyc}
\bibitem{Giacomelli:2023qyc}
S.~Giacomelli and R.~Savelli,
``\ensuremath{\mathcal{N}} = 1 SCFTs from F-theory on Orbifolds,''
\doi{JHEP \textbf{08} (2023), 129}{doi:10.1007/JHEP08(2023)129}
\eprintN{2304.11148}.
%1 citations counted in INSPIRE as of 01 Oct 2023








%\cite{Bianchi:2008cj}
\bibitem{Bianchi:2008cj}
M.~Bianchi,
``Bound-states of D-branes in L-R asymmetric superstring vacua,''
\doi{Nucl. Phys. B \textbf{805} (2008), 168-181}{doi:10.1016/j.nuclphysb.2008.07.008}
\eprintN{0805.3276}.
%11 citations counted in INSPIRE as of 17 Jan 2022
  
  %\cite{Bianchi:2010aw}
\bibitem{Bianchi:2010aw}
M.~Bianchi,
``On $\mathcal{R}^{4}$ terms and MHV amplitudes in $\mathcal{N}$ = 5,6 supergravity vacua of Type II superstrings,''
\doi{Adv. High Energy Phys. \textbf{2011} (2011), 479038}{doi:10.1155/2011/479038}
\eprintN{1010.4736}.
%3 citations counted in INSPIRE as of 22 Mar 2021




%\cite{Bianchi:2022tbr}
\bibitem{Bianchi:2022tbr}
M.~Bianchi, G.~Bossard and D.~Consoli,
``Perturbative higher-derivative terms in $ \mathcal{N} $ = 6 asymmetric orbifolds,''
\doi{JHEP \textbf{06} (2022), 088}{doi:10.1007/JHEP06(2022)088}
\eprintN{2203.15130}.
%2 citations counted in INSPIRE as of 18 Sep 2023



%\cite{Brunner:1999fj}
\bibitem{Brunner:1999fj}
I.~Brunner, A.~Rajaraman and M.~Rozali,
``D-branes on asymmetric orbifolds,''
\doi{Nucl. Phys. B \textbf{558} (1999), 205-215}{doi:10.1016/S0550-3213(99)00376-4}
\eprint{hep-th/9905024}.
%19 citations counted in INSPIRE as of 05 Oct 2023



   %\cite{Gaberdiel:2002jr}
\bibitem{Gaberdiel:2002jr}
M.~R.~Gaberdiel and S.~Schafer-Nameki,
``D-branes in an asymmetric orbifold,''
\doi{Nucl. Phys. B \textbf{654} (2003), 177-196}{doi:10.1016/S0550-3213(03)00062-2}
\eprint{hep-th/0210137}.
%24 citations counted in INSPIRE as of 07 Apr 2021



  %\cite{Kawai:2007qd}
\bibitem{Kawai:2007qd}
S.~Kawai and Y.~Sugawara,
``D-branes in T-fold conformal field theory,''
\doi{JHEP \textbf{02} (2008), 027}{doi:10.1088/1126-6708/2008/02/027}
\eprintN{0709.0257}.
%21 citations counted in INSPIRE as of 07 Apr 2021





%\cite{Bianchi:1990yu}
\bibitem{Bianchi:1990yu}
M.~Bianchi and A.~Sagnotti,
``On the systematics of open string theories,''
\doi{Phys. Lett. B \textbf{247} (1990), 517-524.}{doi:10.1016/0370-2693(90)91894-H}
%551 citations counted in INSPIRE as of 26 Mar 2022

%\cite{Bianchi:1990tb}
\bibitem{Bianchi:1990tb}
M.~Bianchi and A.~Sagnotti,
``Twist symmetry and open string Wilson lines,''
\doi{Nucl. Phys. B \textbf{361} (1991), 519-538.}{doi:10.1016/0550-3213(91)90271-X}
%430 citations counted in INSPIRE as of 26 Mar 2022

%\cite{Harvey:1999gq}
\bibitem{Harvey:1999gq}
J.~A.~Harvey, S.~Kachru, G.~W.~Moore and E.~Silverstein,
``Tension is dimension,''
\doi{JHEP \textbf{03} (2000), 001}{doi:10.1088/1126-6708/2000/03/001}
\eprint{hep-th/9909072}.
%83 citations counted in INSPIRE as of 22 Sep 2023



%\cite{Recknagel:2002qq}
\bibitem{Recknagel:2002qq}
A.~Recknagel,
``Permutation branes,''
\doi{JHEP \textbf{04} (2003), 041}{doi:10.1088/1126-6708/2003/04/041}
\eprint{hep-th/0208119}.
%80 citations counted in INSPIRE as of 22 Sep 2023



%\cite{Dijkgraaf:1989hb}
\bibitem{Dijkgraaf:1989hb}
R.~Dijkgraaf, C.~Vafa, E.~P.~Verlinde and H.~L.~Verlinde,
``The Operator Algebra of Orbifold Models,''
\doi{Commun. Math. Phys. \textbf{123} (1989), 485}{doi:10.1007/BF01238812}
%347 citations counted in INSPIRE as of 20 May 2023





%\cite{Collinucci:2014qfa}
\bibitem{Collinucci:2014qfa}
A.~Collinucci and R.~Savelli,
``T-branes as branes within branes,''
\doi{JHEP \textbf{09} (2015), 161}{doi:10.1007/JHEP09(2015)161}
\eprintN{1410.4178}.
%62 citations counted in INSPIRE as of 01 Oct 2023

%\cite{Bena:2019rth}
\bibitem{Bena:2019rth}
I.~Bena, J.~Bl{a}b\"ack, R.~Savelli and G.~Zoccarato,
``The two faces of T-branes,''
\doi{JHEP \textbf{10} (2019), 150}{doi:10.1007/JHEP10(2019)150}
\eprintN{1905.03267}.
%8 citations counted in INSPIRE as of 01 Oct 2023

%\cite{Lerche:1988np}
\bibitem{Lerche:1988np}
W.~Lerche, A.~N.~Schellekens and N.~P.~Warner,
``Lattices and Strings,''
\doi{Phys. Rept. \textbf{177} (1989), 1}{doi:10.1016/0370-1573(89)90077-X}.
%204 citations counted in INSPIRE as of 16 Oct 2023

%%%%%%%%%%%%%%%%%%%%%%%%%%%%%%%%%%%%%%%%%%%%%%%%%%%%%%%
%%%%%%%%%%%%%%%%%%%%%%%%%%%%%%%%%%%%%%%%%%%%%%%%%%%%%%%
%%%%%%%%%%%%%%%%%%%%%%%%%%%%%%%%%%%%%%%%%%%%%%%%%%%%%%%
%%%%%%%%%%%%%%%%%%%%%%%%%%%%%%%%%%%%%%%%%%%%%%%%%%%%%%%
%%%%%%%%%%%%%%%%%%%%%%%%%%%%%%%%%%%%%%%%%%%%%%%%%%%%%%%
%%%%%%%%%%%%%%%%%%%%%%%%%%%%%%%%%%%%%%%%%%%%%%%%%%%%%%%




%
%
%%\cite{Schnabl:2019oom}
%\bibitem{Schnabl:2019oom}
%M.~Schnabl and J.~Vo\v{s}mera,
%``Gepner-like boundary states on $T^4$,''
%\eprintN{1903.00487}.
%%6 citations counted in INSPIRE as of 05 Oct 2023
%





%282 citations counted in INSPIRE as of 04 Jan 2023


%  
%  %\cite{Cremmer:1978ds}
%\bibitem{Cremmer:1978ds}
%E.~Cremmer and B.~Julia,
%``The N=8 Supergravity Theory 1. The Lagrangian,''
%\doi{Phys. Lett. B \textbf{80} (1978), 48.}{doi:10.1016/0370-2693(78)90303-9}
%%717 citations counted in INSPIRE as of 21 Jan 2022
%
%%\cite{Bianchi:2009wj}
%\bibitem{Bianchi:2009wj}
%M.~Bianchi, S.~Ferrara and R.~Kallosh,
%``Perturbative and non-perturbative $\cN=8$ Supergravity,''
%\doi{Phys. Lett. B \textbf{690} (2010), 328-331}{doi:10.1016/j.physletb.2010.05.049}
%\eprintN{0910.3674}.
%%29 citations counted in INSPIRE as of 10 Mar 2022
%
%%\cite{Bianchi:2009mj}
%\bibitem{Bianchi:2009mj}
%M.~Bianchi, S.~Ferrara and R.~Kallosh,
%``Observations on Arithmetic Invariants and U-Duality Orbits in $\cN =8$ Supergravity,''
%\doi{JHEP \textbf{03} (2010), 081}{doi:10.1007/JHEP03(2010)081}
%\eprintN{0912.0057}.
%%23 citations counted in INSPIRE as of 10 Mar 2022
%
%

%
%\cite{Dabholkar:1998kv}
%\bibitem{Dabholkar:1998kv}
%A.~Dabholkar and J.~A.~Harvey,
%``String islands,''
%\doi{JHEP \textbf{02} (1999), 006}{doi:10.1088/1126-6708/1999/02/006}
%\eprint{hep-th/9809122}.
%
%
%
%%\cite{Andrianopoli:2001zh}
%\bibitem{Andrianopoli:2001zh}
%L.~Andrianopoli, R.~D'Auria and S.~Ferrara,
%``Supersymmetry reduction of N extended supergravities in four-dimensions,''
%\doi{JHEP \textbf{03} (2002), 025}{doi:10.1088/1126-6708/2002/03/025}
%\eprint{hep-th/0110277}.
%%83 citations counted in INSPIRE as of 24 Mar 2022

%\bibitem{Witten:1995ex}
%E.~Witten, ``{String theory dynamics in various dimensions},''
%  \href{http://dx.doi.org/10.1016/0550-3213(95)00158-O}{{\em Nucl.Phys.} {\bf
%  B443} (1995)  85--126},
%\eprint{hep-th/9503124}.
%%%CITATION = HEP-TH/9503124;%%.
%
%%%%%%%%%%%%%%%%%%%%%%%%%%%%%%%%%%%%%%%%%%%%%%%%%%%%%%%%%%%%%%%%%
%
%\bibitem{Green:1981yb}
%M.~B. Green and J.~H. Schwarz, ``{Supersymmetrical String Theories},''
%\href{http://dx.doi.org/10.1016/0370-2693(82)91110-8}{{\em Phys. Lett.} {\bf
%  B109} (1982)  444--448}.
%%%CITATION = PHLTA,B109,444;%%.
%
%
%\bibitem{Green:1997tv}
%M.~B. Green and M.~Gutperle, 
%``{Effects of D-instantons},''
%\doi{Nucl. Phys. B {\bf 498} (1997)  195--227}{doi:10.1016/S0550-3213(97)00269-1}
%\eprint{hep-th/9701093}.
%%%CITATION = HEP-TH/9701093;%%.
%
%\bibitem{Green:1997di}
%M.~B. Green and P.~Vanhove, ``{D-instantons, strings and M-theory},''
%  \href{http://dx.doi.org/10.1016/S0370-2693(97)00785-5}{{\em Phys. Lett.} {\bf
%  B408} (1997)  122--134},
%\eprint{hep-th/9704145}.
%%%CITATION = HEP-TH/9704145;%%.
%
%
%%\cite{Berkovits:1997pj}
%\bibitem{Berkovits:1997pj}
%N.~Berkovits,
%``Construction of $R^4$ terms in $N=2$  $D = 8$ superspace,''
%\doi{Nucl. Phys. B \textbf{514} (1998), 191-203}{doi:10.1016/S0550-3213(97)00817-1}
%\eprint{hep-th/9709116}.
%%46 citations counted in INSPIRE as of 24 Mar 2022
%
%%\cite{Pioline:1998mn}
%\bibitem{Pioline:1998mn}
%B.~Pioline,
%``A note on nonperturbative $R^4$ couplings,''
%\doi{Phys. Lett. B \textbf{431} (1998), 73-76}{doi:10.1016/S0370-2693(98)00554-1}
%\eprint{hep-th/9804023}.
%%76 citations counted in INSPIRE as of 14 Apr 2021
%
%
%%\cite{Green:1998by}
%\bibitem{Green:1998by}
%M.~B.~Green and S.~Sethi,
%``Supersymmetry constraints on type IIB supergravity,''
%\doi{Phys. Rev. D \textbf{59} (1999), 046006}{doi:10.1103/PhysRevD.59.046006}
%\eprintN{hep-th/9808061}.
%%232 citations counted in INSPIRE as of 14 Apr 2021
%
%
%
%%\cite{Obers:1999um}
%\bibitem{Obers:1999um}
%N.~A.~Obers and B.~Pioline,
%``Eisenstein series and string thresholds,''
%\doi{Commun. Math. Phys. \textbf{209} (2000), 275-324}{doi:10.1007/s002200050022}
%\eprint{hep-th/9903113}.
%%105 citations counted in INSPIRE as of 21 Jan 2022
%
%\bibitem{Green:1999pv}
%M.~B. Green and P.~Vanhove, ``{The low energy expansion of the one-loop type II
%  superstring amplitude},''
%  \href{http://dx.doi.org/10.1103/PhysRevD.61.104011}{{\em Phys. Rev.} {\bf
%  D61} (2000)  104011},
%\eprint{hep-th/9910056}.
%%%CITATION = HEP-TH/9910056;%%.
%
%
%
%
%
%%\cite{Obers:2001sw}
%\bibitem{Obers:2001sw}
%N.~A.~Obers and B.~Pioline,
%``Exact thresholds and instanton effects in string theory,''
%Fortsch. Phys. \textbf{49} (2001), 359-375 
%\eprint{hep-th/0101122}.
%%5 citations counted in INSPIRE as of 01 Apr 2021
%
%
%%\cite{Kazhdan:2001nx}
%\bibitem{Kazhdan:2001nx}
%D.~Kazhdan, B.~Pioline and A.~Waldron,
%``Minimal representations, spherical vectors, and exceptional theta series,''
%\doi{Commun. Math. Phys. \textbf{226} (2002), 1-40}{doi:10.1007/s002200200601}
%\eprintN{hep-th/0107222}.
%%46 citations counted in INSPIRE as of 14 Apr 2021
%
%\bibitem{Basu:2008cf}
%A.~Basu and S.~Sethi, ``{Recursion relations from space-time supersymmetry},''
%  \href{http://dx.doi.org/10.1088/1126-6708/2008/09/081}{{\em JHEP} {\bf 09}
%  (2008)  081},
%\eprintN{0808.1250}.
%%%CITATION = ARXIV:0808.1250;%%.
%
%
%
%  %\cite{Green:2005ba}
%\bibitem{Green:2005ba}
%M.~B.~Green and P.~Vanhove,
%``Duality and higher derivative terms in M theory,''
%\doi{JHEP \textbf{01} (2006), 093}{doi:10.1088/1126-6708/2006/01/093}
%\eprintN{hep-th/0510027}.
%%149 citations counted in INSPIRE as of 15 Apr 2021
%
%
%
%
%
%\bibitem{Pioline:2010kb}
%B.~Pioline, ``{$R^4$ couplings and automorphic unipotent representations},''
%  \href{http://dx.doi.org/10.1007/JHEP03(2010)116}{{\em JHEP} {\bf 03} (2010)
%  116},
%\eprintN{1001.3647}.
%%%CITATION = 1001.3647;%%.
%
%
%%\cite{Green:2011vz}
%\bibitem{Green:2011vz}
%M.~B.~Green, S.~D.~Miller and P.~Vanhove,
%``Small representations, string instantons, and Fourier modes of Eisenstein series,''
%\doi{J. Number Theor. \textbf{146} (2015), 187-309}{doi:10.1016/j.jnt.2013.05.018}
%\eprintN{1111.2983}.
%%42 citations counted in INSPIRE as of 14 Apr 2021
%
%
%%\cite{Bossard:2014lra}
%\bibitem{Bossard:2014lra}
%G.~Bossard and V.~Verschinin,
%``Minimal unitary representations from supersymmetry,''
%\doi{JHEP \textbf{10} (2014), 008}{doi:10.1007/JHEP10(2014)008}
%\eprintN{1406.5527}.
%%25 citations counted in INSPIRE as of 14 Apr 2021
%
%%\cite{Bossard:2014aea}
%\bibitem{Bossard:2014aea}
%G.~Bossard and V.~Verschinin,
%``$\mathcal{E} {D}^4 \mathcal{R}^4$ type invariants and their gradient expansion,''
%\doi{JHEP \textbf{03} (2015), 089}{doi:10.1007/JHEP03(2015)089}
%\eprintN{1411.3373}.
%%26 citations counted in INSPIRE as of 14 Apr 2021
%
%
%%\cite{Gustafsson:2014iva}
%\bibitem{Gustafsson:2014iva}
%H.~P.~A.~Gustafsson, A.~Kleinschmidt and D.~Persson,
%``Small automorphic representations and degenerate Whittaker vectors,''
%\doi{J. Number Theor. \textbf{166} (2016), 344-399}{doi:10.1016/j.jnt.2016.02.002}
%\eprintNT{1412.5625}.
%%9 citations counted in INSPIRE as of 24 Mar 2022
%
%
%\bibitem{Wang:2015jna}
%Y.~Wang and X.~Yin, ``{Constraining higher derivative supergravity with
%  scattering amplitudes},''
%  \href{http://dx.doi.org/10.1103/PhysRevD.92.041701}{{\em Phys. Rev.} {\bf
%  D92} (2015) no.~4, 041701},
%\eprintN{1502.03810}.
%%%CITATION = ARXIV:1502.03810;%%.
%
%%\cite{Bossard:2015uga}
%\bibitem{Bossard:2015uga}
%G.~Bossard and V.~Verschinin,
%``The two $\nabla^{6} R^{4}$ type invariants and their higher order generalisation,''
%\doi{JHEP \textbf{07} (2015), 154}{doi:10.1007/JHEP07(2015)154}
%\eprintN{1503.04230}.
%%22 citations counted in INSPIRE as of 14 Apr 2021
%
%%\cite{Gourevitch:2019knu}
%\bibitem{Gourevitch:2019knu}
%D.~Gourevitch, H.~P.~A.~Gustafsson, A.~Kleinschmidt, D.~Persson and S.~Sahi,
%``Fourier coefficients of minimal and next-to-minimal automorphic representations of simply-laced groups,''
%\doi{Can. J. Math. \textbf{74} (2022) no.1, 122-169}{doi:10.4153/S0008414X20000711}
%\eprintNT{1908.08296}.
%%3 citations counted in INSPIRE as of 24 Mar 2022
%
%
%
%%%%%%%%%%%%%%%%%%%%%%%%%%%%%%%%%%%%%%%%%%%%%%%
%
% 
%
%
%%\cite{Green:1997as}
%\bibitem{Green:1997as}
%M.~B.~Green, M.~Gutperle and P.~Vanhove,
%``One loop in eleven-dimensions,''
%\doi{Phys. Lett. B \textbf{409} (1997), 177-184}{doi:10.1016/S0370-2693(97)00931-3}
%\eprint{hep-th/9706175}.
%%306 citations counted in INSPIRE as of 21 Jan 2022
%
%
%
%\bibitem{Kiritsis:1997em}
%E.~Kiritsis and B.~Pioline, ``{On $R^4$ threshold corrections in type IIB
%  string theory and (p,q) string instantons},''
%  \href{http://dx.doi.org/10.1016/S0550-3213(97)00645-7}{{\em Nucl. Phys.} {\bf
%  B508} (1997)  509--534},
%\eprint{hep-th/9707018}.
%%%CITATION = HEP-TH/9707018;%%.
%
%
%
%\bibitem{Pioline:1997pu}
%B.~Pioline and E.~Kiritsis, ``{U-duality and D-brane combinatorics},''
%  \href{http://dx.doi.org/10.1016/S0370-2693(97)01398-1}{{\em Phys. Lett.} {\bf
%  B418} (1998)  61--69},
%\eprint{hep-th/9710078}.
%%%CITATION = HEP-TH/9710078;%%.
%
%
%%\cite{Green:1999pu}
%\bibitem{Green:1999pu}
%M.~B.~Green, H.~h.~Kwon and P.~Vanhove,
%``Two loops in eleven-dimensions,''
%\doi{Phys. Rev. D \textbf{61} (2000), 104010}{doi:10.1103/PhysRevD.61.104010}
%\eprint{hep-th/9910055}.
%%168 citations counted in INSPIRE as of 21 Jan 2022
%
%\bibitem{Basu:2007ru}
%A.~Basu, ``{The $D^4 R^4$ term in type IIB string theory on $T^2$ and U-duality},'' \href{http://dx.doi.org/10.1103/PhysRevD.77.106003}{{\em Phys.
%  Rev.} {\bf D77} (2008)  106003},
%\eprintN{0708.2950}.
%%%CITATION = 0708.2950;%%.
%
%%\cite{Green:2008uj}
%\bibitem{Green:2008uj}
%  M.~B.~Green, J.~G.~Russo and P.~Vanhove,
%  ``Low energy expansion of the four-particle genus-one amplitude in type II superstring theory,''
% \doi{JHEP {\bf 0802} (2008) 020}{doi:10.1088/1126-6708/2008/02/020}
%  \eprintN{0801.0322}.
%  %%CITATION = doi:10.1088/1126-6708/2008/02/020;%%
%  %65 citations counted in INSPIRE as of 11 Jun 2018
%
%
%%\cite{Green:2010wi}
%\bibitem{Green:2010wi}
%M.~B.~Green, J.~G.~Russo and P.~Vanhove,
%``Automorphic properties of low energy string amplitudes in various dimensions,''
%\doi{Phys. Rev. D \textbf{81} (2010), 086008}{doi:10.1103/PhysRevD.81.086008}
%\eprintN{1001.2535}.
%%85 citations counted in INSPIRE as of 19 Jan 2022
%
%\bibitem{Green:2010sp}
%M.~B. Green, J.~G. Russo, and P.~Vanhove, ``{String theory dualities and
%  supergravity divergences},''
%  \href{http://dx.doi.org/10.1007/JHEP06(2010)075}{{\em JHEP} {\bf 1006} (2010)
%   075},
%\eprintN{1002.3805}.
%%%CITATION = ARXIV:1002.3805;%%.
%
%%\cite{Green:2010kv}
%\bibitem{Green:2010kv}
%M.~B.~Green, S.~D.~Miller, J.~G.~Russo and P.~Vanhove,
%``Eisenstein series for higher-rank groups and string theory amplitudes,''
%\doi{Commun. Num. Theor. Phys. \textbf{4} (2010), 551-596}{doi:10.4310/CNTP.2010.v4.n3.a2}
%\eprintN{1004.0163}.
%%77 citations counted in INSPIRE as of 21 Jan 2022
%
%%\cite{Green:2014yxa}
%\bibitem{Green:2014yxa}
%M.~B.~Green, S.~D.~Miller and P.~Vanhove,
%``$SL(2, \mathbb{Z})$-invariance and D-instanton contributions to the $D^6 R^4$ interaction,''
%\doi{Commun. Num. Theor. Phys. \textbf{09} (2015), 307-344}{doi:10.4310/CNTP.2015.v9.n2.a3}
%\eprintN{1404.2192}.
%%45 citations counted in INSPIRE as of 21 Jan 2022
%
%
%
%%\cite{Bossard:2020xod}
%\bibitem{Bossard:2020xod}
%G.~Bossard, A.~Kleinschmidt and B.~Pioline,
%``1/8-BPS Couplings and Exceptional Automorphic Functions,''
%\doi{SciPost Phys. \textbf{8} (2020) no.4, 054}{doi:10.21468/SciPostPhys.8.4.054}
%\eprintN{2001.05562}.
%%5 citations counted in INSPIRE as of 21 Jan 2022
%
%
%  %\cite{Harvey:1996ir}
%\bibitem{Harvey:1996ir}
%J.~A.~Harvey and G.~W.~Moore,
%``Five-brane instantons and $R^2$ couplings in $\mathcal{N}=4$ string theory,''
%\doi{Phys. Rev. D \textbf{57} (1998), 2323-2328}{doi:10.1103/PhysRevD.57.2323}
%\eprint{hep-th/9610237}.
%%108 citations counted in INSPIRE as of 05 Mar 2022
%
%
%
%%\cite{Bachas:1997mc}
%\bibitem{Bachas:1997mc}
%C.~Bachas, C.~Fabre, E.~Kiritsis, N.~A.~Obers and P.~Vanhove,
%``Heterotic / type I duality and D-brane instantons,''
%\doi{Nucl. Phys. B \textbf{509} (1998), 33-52}{doi:10.1016/S0550-3213(97)00639-1}
%\eprint{hep-th/9707126}.
%%134 citations counted in INSPIRE as of 01 Apr 2021
%
%
%%\cite{Gregori:1997hi}
%\bibitem{Gregori:1997hi}
%A.~Gregori, E.~Kiritsis, C.~Kounnas, N.~A.~Obers, P.~M.~Petropoulos and B.~Pioline,
%``$R^2$ corrections and nonperturbative dualities of N=4 string ground states,''
%\doi{Nucl. Phys. B \textbf{510} (1998), 423-476}{doi:10.1016/S0550-3213(97)00635-4}
%\eprint{hep-th/9708062}.
%%138 citations counted in INSPIRE as of 05 Mar 2022
%
%%\cite{Kiritsis:1997hf}
%\bibitem{Kiritsis:1997hf}
%E.~Kiritsis and N.~A.~Obers,
%``Heterotic type I duality in $D<10$-dimensions, threshold corrections and D instantons,''
%\doi{JHEP \textbf{10} (1997), 004}{doi:10.1088/1126-6708/1997/10/004}
%\eprint{hep-th/9709058}.
%%93 citations counted in INSPIRE as of 01 Apr 2021
%
%
%%\cite{Bianchi:1998vq}
%\bibitem{Bianchi:1998vq}
%M.~Bianchi, E.~Gava, J.~F.~Morales and K.~S.~Narain,
%``D strings in unconventional type I vacuum configurations,''
%\doi{Nucl. Phys. B \textbf{547} (1999), 96-126}{doi:10.1016/S0550-3213(99)00004-8}
%\eprint{hep-th/9811013}.
%%33 citations counted in INSPIRE as of 01 Apr 2021
%
%%
%%%\cite{Bianchi:2007rb}
%%\bibitem{Bianchi:2007rb}
%%M.~Bianchi and J.~F.~Morales,
%%``Unoriented D-brane Instantons vs Heterotic worldsheet Instantons,''
%%\doi{JHEP \textbf{02} (2008), 073}{doi:10.1088/1126-6708/2008/02/073}
%%\eprintN{0712.1895}.
%%%33 citations counted in INSPIRE as of 01 Apr 2021
%
%
%  %\cite{Bossard:2018rlt}
%\bibitem{Bossard:2018rlt}
%G.~Bossard, C.~Cosnier-Horeau and B.~Pioline,
%``Exact effective interactions and 1/4-BPS dyons in heterotic CHL orbifolds,''
%\doi{SciPost Phys. \textbf{7} (2019) no.3, 028}{doi:10.21468/SciPostPhys.7.3.028}
%\eprintN{1806.03330}.
%%7 citations counted in INSPIRE as of 19 Apr 2021
%
%\bibitem{Hull:2004in}
%C.~M.~Hull,
%``A Geometry for non-geometric string backgrounds,''
%\doi{JHEP \textbf{10} (2005), 065}{doi:10.1088/1126-6708/2005/10/065}
%\eprint{hep-th/0406102}.
%%524 citations counted in INSPIRE as of 24 Mar 2022
%
%%%%%%%%%%%%%%%%%%%%%%%%%%%%%%%%%%%%%%%%%%%%%%%%%%%%%%%%%%%%%
%%%%%%%%%%%%%%%%%%%%%%%%%%%%%%%%%%%%%%%%%%%%%%%%%%%%%%%%%%%%%
%
%  %\cite{Ferrara:1989nm}
%\bibitem{Ferrara:1989nm}
%S.~Ferrara and C.~Kounnas,
%``Extended supersymmetry in four-dimensional Type {II} Strings,''
%\doi{Nucl. Phys. B \textbf{328} (1989), 406-438.}{doi:10.1016/0550-3213(89)90335-0}
%%67 citations counted in INSPIRE as of 17 Jan 2022
%
%%\cite{Bianchi:2008cj}
%\bibitem{Bianchi:2008cj}
%M.~Bianchi,
%``Bound-states of D-branes in L-R asymmetric superstring vacua,''
%\doi{Nucl. Phys. B \textbf{805} (2008), 168-181}{doi:10.1016/j.nuclphysb.2008.07.008}
%\eprintN{0805.3276}.
%%11 citations counted in INSPIRE as of 17 Jan 2022
%  
%  %\cite{Bianchi:2010aw}
%\bibitem{Bianchi:2010aw}
%M.~Bianchi,
%``On $\mathcal{R}^{4}$ terms and MHV amplitudes in $\mathcal{N}$ = 5,6 supergravity vacua of Type II superstrings,''
%\doi{Adv. High Energy Phys. \textbf{2011} (2011), 479038}{doi:10.1155/2011/479038}
%\eprintN{1010.4736}.
%%3 citations counted in INSPIRE as of 22 Mar 2021
%
%%\cite{Narain:1986am}
%\bibitem{Narain:1986am}
%K.~S.~Narain, M.~H.~Sarmadi and E.~Witten,
%``A note on toroidal compactification of Heterotic String Theory,''
%\doi{Nucl. Phys. B \textbf{279} (1987), 369-379.}{doi:10.1016/0550-3213(87)90001-0}
%%790 citations counted in INSPIRE as of 24 Mar 2022
%
%
%
%
%
%
%%\cite{Dabholkar:1998kv}
%\bibitem{Dabholkar:1998kv}
%A.~Dabholkar and J.~A.~Harvey,
%``String islands,''
%\doi{JHEP \textbf{02} (1999), 006}{doi:10.1088/1126-6708/1999/02/006}
%\eprint{hep-th/9809122}.
%%61 citations counted in INSPIRE as of 13 Jan 2022
%
%%\cite{Gunaydin:1983bi}
%\bibitem{Gunaydin:1983bi}
%M.~G\"{u}naydin, G.~Sierra and P.~K.~Townsend,
%``The Geometry of N=2 Maxwell--Einstein Supergravity and Jordan Algebras,''
%\doi{Nucl. Phys. B \textbf{242} (1984), 244-268.}{doi:10.1016/0550-3213(84)90142-1}
%%581 citations counted in INSPIRE as of 01 Apr 2021
%
%%\cite{Bianchi:2007va}
%\bibitem{Bianchi:2007va}
%M.~Bianchi and S.~Ferrara,
%``Enriques and Octonionic Magic Supergravity Models,''
%\doi{JHEP \textbf{02} (2008), 054}{doi:10.1088/1126-6708/2008/02/054}
%\eprintN{0712.2976}.
%%22 citations counted in INSPIRE as of 10 Mar 2022
%
%  
%\bibitem{Krieg}
%A.~Krieg, ``Modular forms on half-spaces of quaternions'', Lecture Notes  in Mathematics {\bf 1143} (1985).
% 
%
%\bibitem{Vigneras}
%Marie-France Vign\'eras , “Arithm\'etique des alg\`ebres de quaternions'' Lecture Notes in Mathematics  (1980) 800.
%
%
%
%\bibitem{GreenSchwWitt}
%M.B. Green, J.H. Schwarz and E. Witten, ``Superstring theory,'' Cambridge University Press.
%
%
%%\cite{Bossard:2010bd}
%\bibitem{Bossard:2010bd}
%G.~Bossard, P.~S.~Howe and K.~S.~Stelle,
%``On duality symmetries of supergravity invariants,''
%\doi{JHEP \textbf{01} (2011), 020}{doi:10.1007/JHEP01(2011)020}
%\eprintN{1009.0743}.
%%86 citations counted in INSPIRE as of 21 Jan 2022
%
%%\cite{Bern:2013uka}
%\bibitem{Bern:2013uka}
%Z.~Bern, S.~Davies, T.~Dennen, A.~V.~Smirnov and V.~A.~Smirnov,
%``Ultraviolet properties of $\mathcal{N}=4$ Supergravity at four loops,''
%\doi{Phys. Rev. Lett. \textbf{111} (2013) no.23, 231302}{doi:10.1103/PhysRevLett.111.231302}
%\eprintN{1309.2498}.
%%142 citations counted in INSPIRE as of 21 Jan 2022
%
%
%%\cite{Bern:2014sna}
%\bibitem{Bern:2014sna}
%Z.~Bern, S.~Davies and T.~Dennen,
%``Enhanced ultraviolet cancellations in $\mathcal N=5$ supergravity at four loops,''
%\doi{Phys. Rev. D \textbf{90} (2014) no.10, 105011}{doi:10.1103/PhysRevD.90.105011}
%\eprintN{1409.3089}.
%%144 citations counted in INSPIRE as of 21 Jan 2022
%
%
%
%%\cite{Bossard:2011tq}
%\bibitem{Bossard:2011tq}
%G.~Bossard, P.~S.~Howe, K.~S.~Stelle and P.~Vanhove,
%``The vanishing volume of D=4 superspace,''
%\doi{Class. Quant. Grav. \textbf{28} (2011), 215005}{doi:10.1088/0264-9381/28/21/215005}
%\eprintN{1105.6087}.
%%82 citations counted in INSPIRE as of 21 Jan 2022
%
%
%
%%\cite{Bianchi:2015vsa}
%\bibitem{Bianchi:2015vsa}
%M.~Bianchi and D.~Consoli,
%``Simplifying one-loop amplitudes in superstring theory,''
%\doi{JHEP \textbf{01} (2016), 043}{doi:10.1007/JHEP01(2016)043}
%\eprintN{1508.00421}.
%%17 citations counted in INSPIRE as of 13 Jan 2022
%
%
%%\cite{Bianchi:2010mg}
%\bibitem{Bianchi:2010mg}
%M.~Bianchi, R.~Poghossian and M.~Samsonyan,
%``Precision Spectroscopy and Higher Spin symmetry in the ABJM model,''
%\doi{JHEP \textbf{10} (2010), 021}{doi:10.1007/JHEP10(2010)021}
%\eprintN{1005.5307}.
%%14 citations counted in INSPIRE as of 10 Mar 2022
%
%
%
%%\cite{Kiritsis:1997hj}
%\bibitem{Kiritsis:1997hj}
%E.~Kiritsis,
%``Introduction to superstring theory,''
%\eprint{hep-th/9709062}.
%%223 citations counted in INSPIRE as of 21 Jan 2022
%
%
%
%
%%\cite{Bergshoeff:1997gy}
%\bibitem{Bergshoeff:1997gy}
%E.~Bergshoeff, B.~Janssen and T.~Ort\'{\i}n,
%``Kaluza-Klein monopoles and gauged sigma models,''
%\doi{Phys. Lett. B \textbf{410} (1997), 131-141}{doi:10.1016/S0370-2693(97)00946-5}
%\eprint{hep-th/9706117}.
%%95 citations counted in INSPIRE as of 24 Feb 2022
%
%
%
%%\cite{Alvarez-Gaume:1983ihn}
%\bibitem{Alvarez-Gaume:1983ihn}
%L.~\'{A}lvarez-Gaum\'{e} and E.~Witten,
%``Gravitational Anomalies,''
%\doi{Nucl. Phys. B \textbf{234} (1984), 269.}{doi:10.1016/0550-3213(84)90066-X}
%%1712 citations counted in INSPIRE as of 15 Mar 2022
%
%
%
%
%%\cite{Andrianopoli:2001zh}
%\bibitem{Andrianopoli:2001zh}
%L.~Andrianopoli, R.~D'Auria and S.~Ferrara,
%``Supersymmetry reduction of N extended supergravities in four-dimensions,''
%\doi{JHEP \textbf{03} (2002), 025}{doi:10.1088/1126-6708/2002/03/025}
%\eprint{hep-th/0110277}.
%%83 citations counted in INSPIRE as of 24 Mar 2022
%
%%\cite{Narain:1986qm}
%\bibitem{Narain:1986qm}
%K.~S.~Narain, M.~H.~Sarmadi and C.~Vafa,
%``Asymmetric Orbifolds,''
%\doi{Nucl. Phys. B \textbf{288} (1987), 551.}{doi:10.1016/0550-3213(87)90228-8}
%%489 citations counted in INSPIRE as of 24 Mar 2022
%
%
%%\cite{Narain:1990mw}
%\bibitem{Narain:1990mw}
%K.~S.~Narain, M.~H.~Sarmadi and C.~Vafa,
%``Asymmetric orbifolds: Path integral and operator formulations,''
%\doi{Nucl. Phys. B \textbf{356} (1991), 163-207.}{doi:10.1016/0550-3213(91)90145-N}
%%88 citations counted in INSPIRE as of 24 Mar 2022
%
%
%
%%\cite{Anastasopoulos:2009kj}
%\bibitem{Anastasopoulos:2009kj}
%P.~Anastasopoulos, M.~Bianchi, J.~F.~Morales and G.~Pradisi,
%``(Unoriented) T-folds with few T's,''
%\doi{JHEP \textbf{06} (2009), 032}{doi:10.1088/1126-6708/2009/06/032}
%\eprintN{0901.0113}.
%%10 citations counted in INSPIRE as of 26 Mar 2022
%

%

%
%
%
%
%%\cite{Andrianopoli:2006ub}
%\bibitem{Andrianopoli:2006ub}
%L.~Andrianopoli, R.~D'Auria, S.~Ferrara and M.~Trigiante,
%``Extremal black holes in supergravity,''
%Lect. Notes Phys. \textbf{737} (2008), 661-727, 
%\eprint{hep-th/0611345}.
%%131 citations counted in INSPIRE as of 24 Mar 2022
%
%
%
%
%
%%\cite{Persson:2015jka}
%\bibitem{Persson:2015jka}
%D.~Persson and R.~Volpato,
%``Fricke S-duality in CHL models,''
%\doi{JHEP \textbf{12} (2015), 156}{doi:10.1007/JHEP12(2015)156}
%\eprintN{1504.07260}.
%%28 citations counted in INSPIRE as of 09 Mar 2022
%
%
%\bibitem{GBinprep}
%G.~Bossard, in preparation.
%
%
%
%%\cite{Cremmer:1979uq}
%\bibitem{Cremmer:1979uq}
%E.~Cremmer, J.~Scherk and J.~H.~Schwarz,
%``Spontaneously broken $N=8$ supergravity,''
%\doi{Phys. Lett. B \textbf{84} (1979), 83-86.}{doi:10.1016/0370-2693(79)90654-3}
%%271 citations counted in INSPIRE as of 09 Mar 2022
%
%
%
%
% %\cite{Drummond:2003ex}
%\bibitem{Drummond:2003ex}
%J.~M.~Drummond, P.~J.~Heslop, P.~S.~Howe and S.~F.~Kerstan,
%``Integral invariants in N=4 SYM and the effective action for coincident D-branes,''
%\doi{JHEP \textbf{08} (2003), 016}{doi:10.1088/1126-6708/2003/08/016}
%\eprint{hep-th/0305202}.
%%76 citations counted in INSPIRE as of 14 Apr 2021
%
%
%
%%\cite{Drummond:2010fp}
%\bibitem{Drummond:2010fp}
%J.~M.~Drummond, P.~J.~Heslop and P.~S.~Howe,
%``A Note on N=8 counterterms,''
%\eprintN{1008.4939}.
%%26 citations counted in INSPIRE as of 14 Apr 2021
%  
%  
%  
%%\cite{Howe:1981gz}
%\bibitem{Howe:1981gz}
%P.~S.~Howe,
%``Supergravity in Superspace,''
%\doi{Nucl. Phys. B \textbf{199} (1982), 309-364.}{doi:10.1016/0550-3213(82)90349-2}
%%174 citations counted in INSPIRE as of 24 Mar 2022
%
%
%%\cite{Dobrev:1985qv}
%\bibitem{Dobrev:1985qv}
%V.~K.~Dobrev and V.~B.~Petkova,
%``All positive energy unitary irreducible representations of extended conformal supersymmetry,''
%\doi{Phys. Lett. B \textbf{162} (1985), 127-132.}{doi:10.1016/0370-2693(85)91073-1}
%%311 citations counted in INSPIRE as of 15 Apr 2021
%
%  
%  
%%\cite{Ferrara:1999zg}
%\bibitem{Ferrara:1999zg}
%S.~Ferrara and E.~Sokatchev,
%``Short representations of $SU(2,2|N)$ and harmonic superspace analyticity,''
%\doi{Lett. Math. Phys. \textbf{52} (2000), 247-262}{doi:10.1023/A:1007641619266}
%\eprint{hep-th/9912168}.
%%47 citations counted in INSPIRE as of 14 Apr 2021
%
%
%  \bibitem{CollingwoodMcGovern}
%D.~H. Collingwood and W.~M. McGovern, {\em Nilpotent orbits in semisimple {L}ie
%  algebras}.
%\newblock Van Nostrand Reinhold Mathematics Series. Van Nostrand Reinhold Co.,
%  New York, 1993.
%
%
%
%
%
%
%
%  %\cite{DHoker:2014oxd}
%\bibitem{DHoker:2014oxd}
%E.~D'Hoker, M.~B.~Green, B.~Pioline and R.~Russo,
%``Matching the $D^{6}R^{4}$ interaction at two-loops,''
%\doi{JHEP \textbf{01} (2015), 031}{doi:10.1007/JHEP01(2015)031}
%\eprintN{1405.6226}.
%%64 citations counted in INSPIRE as of 07 Oct 2021
%
%
%
%%\cite{Berg:2016wux}
%\bibitem{Berg:2016wux}
%M.~Berg, I.~Buchberger and O.~Schlotterer,
%``From maximal to minimal supersymmetry in string loop amplitudes,''
%\doi{JHEP \textbf{04} (2017), 163}{doi:10.1007/JHEP04(2017)163}
%\eprintN{1603.05262}.
%%25 citations counted in INSPIRE as of 24 Feb 2022
%
%
%
%%\cite{Bianchi:2012ud}
%\bibitem{Bianchi:2012ud}
%M.~Bianchi and G.~Inverso,
%``Unoriented D-brane instantons,''
%\doi{Fortsch. Phys. \textbf{60} (2012), 822-834}{doi:10.1002/prop.201200047}
%\eprintN{1202.6508}.
%%15 citations counted in INSPIRE as of 24 Mar 2022
%
%%\cite{Bianchi:2012kt}
%\bibitem{Bianchi:2012kt}
%M.~Bianchi, G.~Inverso and L.~Martucci,
%``Brane instantons and fluxes in F-theory,''
%\doi{JHEP \textbf{07} (2013), 037}{doi:10.1007/JHEP07(2013)037}
%\eprintN{1212.0024}.
%%26 citations counted in INSPIRE as of 24 Mar 2022
%
%
%
%%%%%%%%%%%%%%%%%%%%%%%%%%%%%%%%%%%%%%%%%%%%%%%%%%%%%%%
%
%%\cite{Kawai:1985xq}
%\bibitem{Kawai:1985xq}
%H.~Kawai, D.~C.~Lewellen and S.~H.~H.~Tye,
%``A Relation Between Tree Amplitudes of Closed and Open Strings,''
%\doi{Nucl. Phys. B \textbf{269} (1986), 1-23}{doi:10.1016/0550-3213(86)90362-7}
%%986 citations counted in INSPIRE as of 24 Mar 2022
%
%
%
%%\cite{Bianchi:2006nf}
%\bibitem{Bianchi:2006nf}
%M.~Bianchi and A.~V.~Santini,
%``String predictions for near future colliders from one-loop scattering amplitudes around D-brane worlds,''
%\doi{JHEP \textbf{12} (2006), 010}{doi:10.1088/1126-6708/2006/12/010}
%\eprint{hep-th/0607224}.
%%40 citations counted in INSPIRE as of 31 Mar 2021
%
%%\cite{Bern:2011rj}
%\bibitem{Bern:2011rj}
%Z.~Bern, C.~Boucher-Veronneau and H.~Johansson,
%``$\mathcal{N} \ge  4$ supergravity amplitudes from gauge theory at one loop,''
%\doi{Phys. Rev. D \textbf{84} (2011), 105035}{doi:10.1103/PhysRevD.84.105035}
%\eprintN{1107.1935}.
%%113 citations counted in INSPIRE as of 27 Sep 2021
%
%
%%\cite{David:2006ud}
%\bibitem{David:2006ud}
%J.~R.~David, D.~P.~Jatkar and A.~Sen,
%``Dyon spectrum in generic $\mathcal{N}=4$ supersymmetric $\mathds{Z}_N$ orbifolds,''
%\doi{JHEP \textbf{01} (2007), 016}{doi:10.1088/1126-6708/2007/01/016}
%\eprint{hep-th/0609109}.
%%84 citations counted in INSPIRE as of 05 Mar 2022
%
%%\cite{Bachas:2008jv}
%\bibitem{Bachas:2008jv}
%C.~Bachas, M.~Bianchi, R.~Blumenhagen, D.~L\"ust and T.~Weigand,
%``Comments on Orientifolds without Vector Structure,''
%\doi{JHEP \textbf{08} (2008), 016}{doi:10.1088/1126-6708/2008/08/016}
%\eprintN{0805.3696}.
%%25 citations counted in INSPIRE as of 10 Mar 2022
%
%
%%\cite{Bossard:2016hgy}
%\bibitem{Bossard:2016hgy}
%G.~Bossard and B.~Pioline,
%``Exact $\nabla^4 R^4$ couplings and helicity supertraces,''
%\doi{JHEP \textbf{01} (2017), 050}{doi:10.1007/JHEP01(2017)050}
%\eprintN{1610.06693}.
%%10 citations counted in INSPIRE as of 09 Apr 2021
%
%
%   %\cite{Gaberdiel:2002jr}
%\bibitem{Gaberdiel:2002jr}
%M.~R.~Gaberdiel and S.~Schafer-Nameki,
%``D-branes in an asymmetric orbifold,''
%\doi{Nucl. Phys. B \textbf{654} (2003), 177-196}{doi:10.1016/S0550-3213(03)00062-2}
%\eprint{hep-th/0210137}.
%%24 citations counted in INSPIRE as of 07 Apr 2021
%
%  %\cite{Kawai:2007qd}
%\bibitem{Kawai:2007qd}
%S.~Kawai and Y.~Sugawara,
%``D-branes in T-fold conformal field theory,''
%\doi{JHEP \textbf{02} (2008), 027}{doi:10.1088/1126-6708/2008/02/027}
%\eprintN{0709.0257}.
%%21 citations counted in INSPIRE as of 07 Apr 2021
%
%%\cite{Bianchi:2016bgx}
%\bibitem{Bianchi:2016bgx}
%M.~Bianchi, J.~F.~Morales and L.~Pieri,
%``Stringy origin of $4D$ black hole microstates,''
%\doi{JHEP \textbf{06} (2016), 003}{doi:10.1007/JHEP06(2016)003}
%\eprintN{1603.05169}.
%%18 citations counted in INSPIRE as of 11 Mar 2022
%
%%\cite{Bianchi:2017sds}
%\bibitem{Bianchi:2017sds}
%M.~Bianchi, D.~Consoli and J.~F.~Morales,
%``Probing Fuzzballs with Particles, Waves and Strings,''
%\doi{JHEP \textbf{06} (2018), 157}{doi:10.1007/JHEP06(2018)157}
%\eprintN{1711.10287}.
%%21 citations counted in INSPIRE as of 11 Mar 2022
%
%
%%\cite{Anastasopoulos:2011hj}
%\bibitem{Anastasopoulos:2011hj}
%P.~Anastasopoulos, M.~Bianchi and R.~Richter,
%``Light stringy states,''
%\doi{JHEP \textbf{03} (2012), 068}{doi:10.1007/JHEP03(2012)068}
%\eprintN{1110.5424}.
%%26 citations counted in INSPIRE as of 10 Mar 2022
%
%%\cite{Gava:1997jt}
%\bibitem{Gava:1997jt}
%E.~Gava, K.~S.~Narain and M.~H.~Sarmadi,
%``On the bound states of $p$-branes and $(p+2)$-branes,''
%\doi{Nucl. Phys. B \textbf{504} (1997), 214-238}{doi:10.1016/S0550-3213(97)00508-7}
%\eprint{hep-th/9704006}.
%%123 citations counted in INSPIRE as of 24 Mar 2022
%
%
%%\cite{Angelantonj:2011hs}
%\bibitem{Angelantonj:2011hs}
%C.~Angelantonj, C.~Condeescu, E.~Dudas and G.~Pradisi,
%``Non-perturbative transitions among intersecting-brane vacua,''
%\doi{JHEP \textbf{07} (2011), 123}{doi:10.1007/JHEP07(2011)123}
%\eprintN{1105.3465}.
%%10 citations counted in INSPIRE as of 20 Mar 2022
%
%
%
%\bibitem{FNF}
%Dennis R. Estes and Gordon Nipp, “Factorization in quaternion orders'' Journal of Number Theory {\bf 33}, 224 (1989). 
%
%%\cite{DHoker:2015gmr}
%\bibitem{DHoker:2015gmr}
%E.~D'Hoker, M.~B.~Green and P.~Vanhove,
%``On the modular structure of the genus-one Type II superstring low energy expansion,''
%\doi{JHEP \textbf{08} (2015), 041}{doi:10.1007/JHEP08(2015)041}
%\eprintN{1502.06698}.
%%74 citations counted in INSPIRE as of 25 Mar 2022
%
% \bibitem{Weiertheta}
%F. W. J. Olver et al., “NIST Handbook of Mathematical Functions'' Cambridge University
%Press (2010); {{\hypersetup{urlcolor=darkred}\href{http://dlmf.nist.gov}{DLMF, Digital Library of Mathematical Functions}\hypersetup{urlcolor=blue}}}.
%
%
%\bibitem{MR656029}
%D.~Zagier, ``The {R}ankin-{S}elberg method for automorphic functions which are
%  not of rapid decay,'' {\em J. Fac. Sci. Univ. Tokyo Sect. IA Math.} {\bf 28}
%  (1981) no.~3, 415--437 (1982).  
%
%%\cite{Tourkine:2013rda}
%\bibitem{Tourkine:2013rda}
%P.~Tourkine,
%``Tropical Amplitudes,''
%\doi{Annales Henri Poincar\'e \textbf{18} (2017) no.6, 2199-2249}{doi:10.1007/s00023-017-0560-7}
%\eprintN{1309.3551}.
%%36 citations counted in INSPIRE as of 17 Oct 2021
%
\end{thebibliography}
\end{document}